\documentclass[prb,superscriptaddress,twocolumn,floatfix,letterpaper]{revtex4}
\pdfoutput=1

\usepackage{amssymb, amsmath, mathrsfs}
\usepackage[usenames]{xcolor}
\usepackage{subfigure, wrapfig}
\usepackage{threeparttable}
\usepackage{enumerate}
\usepackage[pdftex]{graphicx}
\usepackage[normalem]{ulem}

\definecolor{lightblue}{rgb}{0.4,0.4,1}

\newcommand{\punct}[1]{\textrm{ #1}}
\newcommand{\txt}{\textnormal}

\ifx\officialQ\undefined
	\newcommand{\remark}[1]{{\color{lightblue} (#1)}}

	\newcommand{\deleted}[1]{{\color{brown} \sout{#1}}}
\else
	\newcommand{\remark}[1]{}

	\newcommand{\deleted}[1]{}
\fi

\newcommand{\vv}[1]{\mathbf{#1}}
\newcommand{\vvsym}[1]{\boldsymbol{#1}}
\newcommand{\uv}[1]{\hat{\mathbf{#1}}}

\newcommand{\pd}{\partial}

\newcommand{\up}{\uparrow}
\newcommand{\dn}{\downarrow}
\newcommand{\dg}{\dagger}
\newcommand{\ndg}{\phantom{\dagger}}
\newcommand{\ket}[1]{| #1 \rangle}
\newcommand{\bra}[1]{\langle #1 |}

\newcommand{\avg}[1]{\left\langle #1 \right\rangle}
\newcommand{\avgnormal}[1]{\langle #1 \rangle}

\newcommand{\Proj}{\mathcal{P}}
\newcommand{\E}{\mathcal{E}}
\newcommand{\J}{\mathcal{J}}
\newcommand{\N}{N}
\newcommand{\s}{\mathfrak{s}}
\newcommand{\Hilbert}{\mathscr{H}}
\newcommand{\omitted}{(\ldots)}
\newcommand{\Lg}{\mathcal{L}}
\newcommand{\KleinProd}{\hat{\Gamma}}

\usepackage[pdftex]{hyperref}

\usepackage[pdftex]{hyperref}

\begin{document}

\title{Ordering and criticality in an underscreened Kondo chain }

\author{Wing-Ho Ko}
\author{Hong-Chen Jiang}
\affiliation{Kavli Institute for Theoretical Physics, University of California, Santa Barbara, Santa Barbara, California 93106, USA}
\author{Jeffrey G. Rau}
\affiliation{Department of Physics, University of Toronto, Toronto, Ontario M5S 1A7, Canada}
\author{Leon Balents}
\affiliation{Kavli Institute for Theoretical Physics, University of California, Santa Barbara, Santa Barbara, California 93106, USA}

\date{January 15, 2013}

\begin{abstract} 
Motivated by the nickel valence controversy in the perovskite nickelate \textit{R}NiO$_3$, we consider a one-dimensional underscreened Kondo chain consisting of alternating spin-1 (``nickel'') and electron (``oxygen'') sites, which in addition to the usual electron hopping and spin-spin interaction between the $S=1$ spin and the electron also contains a spin mediated electron hopping term. Using the density-matrix renormalization group (DMRG), we obtained the zero temperature phase diagram of the model, as well as various correlation functions in each phase. Importantly, for a certain range of parameters the model exhibits a quasi-long-range spiral (QS) order. To understand the DMRG results, we construct a mean-field theory based on a Schwinger fermion decomposition of the $S=1$ spins, from which we argue that the QS phase corresponds to a phase in proximity to the spin Bose metal state proposed by D.\@ N.\@ Sheng, O.\@ I.\@ Motrunich, and M.\@ P.\@ A.\@ Fisher [Phys.\@ Rev.\@ B \textbf{79}, 205112 (2009)].  Notably, we find no evidence for a phase with the symmetry of ``nickel''-centered charge order, which has been argued to arise due to site-selective Kondo screening of half the $S=1$ spins, and suggest that order of this type occurs only due to an additional energy gain from spontaneous lattice distortions, not present in this model.  
\end{abstract}

\maketitle 

\section{Introduction} \label{sec:intro}

Mott metal-insulator transitions and their associated spin and charge ordering appear throughout the physics of transition metal oxides, and present many puzzles of interpretation and modeling. Since electrons in the regime of a Mott transition may possess both localized and itinerant character, ambiguity may arise as to an appropriate model Hamiltonian: Hubbard, Heisenberg, or $t$-$J$? Anderson or Kondo model? In multi-orbital systems, the choices are even more numerous. Such a situation arises in the pseudo-cubic perovskite nickelates, \textit{R}NiO$_3$, where \textit{R} is a rare earth ion, which have been studied for many years for their interesting Mott metal-insulator transitions.\cite{imada1998metal} A $T>0$ transition occurs to an insulating state for all members of the series save the case \textit{R}=La, which is metallic at all temperatures, to a low temperature state with complex structural symmetry breaking---denoted ``charge order'' in the literature---and antiferromagnetism. Nominally, in this material the nickel valence is $3+$, i.e. $3d^7$, which would suggest a model with a single electron in the cubic $e_g$ doublet, a two-fold orbital degeneracy. However, another interpretation is that the nickel valence is actually $2+$, with an additional hole per nickel spread out among the oxygen ions,\cite{Mizokawa:PRB:2000} analogous to the Zhang-Rice singlet in cuprates.\cite{FCZhang:PRB:1988} In this ``ligand hole'' scenario there is no orbital degeneracy, but rather an $S=1$ spin on the nickel site, corresponding to a half-filled $e_g$ doublet. The difference between these two scenarios (which we termed the ``nickel valence controversy'') is well defined only in the ionic limit where the charge on the nickel site does not fluctuate significantly, which might be the case in the ligand-hole picture. A third, compromise view is that the valence fluctuates significantly due to hybridization, making the distinction between Ni$^{2+}$ and Ni$^{3+}$ moot; in this case one might build a theory in terms of the low energy bands near the Fermi energy. Though such a compromise exists, it seems to be unpopular, and strong opinions in favor of the ligand-hole picture are often voiced.\cite{park2012site} Going further, Sawatzky has proposed a picture for the observed ``charge'' ordering in terms of local Kondo screening of half the Ni$^{2+}$ spins.\cite{sawatzky}  A recent DMFT paper seems to suggest a similar scenario.\cite{park2012site} A physical understanding of \emph{how} such Kondo screening might come about, and whether it can account for the observed charge/spin order, is, however, still lacking.  

In this paper, we study a concrete model motivated by the ligand-hole scenario  using the numerically exact density matrix renormalization group (DMRG) method.\cite{White:PRL:1992} We write down the simplest model incorporating a sharply defined Ni$^{3+}$ valence, and holes on the oxygen sites. Because each nickel thereby has an $S=1$ spin and is accompanied by only a single oxygen hole with $S=1/2$, this takes the form of an underscreened Kondo lattice model.\cite{Andrei:PRB:2000, Perkins:PRB:2007} To enable the DMRG analysis, we take this model to be one dimensional. Though this is obviously a drastic approximation, our model retains the underscreened Kondo physics, and has the symmetries to allow charge ordering of the same type observed in the nickelate materials. We obtain the numerically exact full zero temperature phase diagram for this model, which contains several magnetic states. However, we \emph{do not} find the charge ordered state seen in the nickelates, which is nickel site-centered, but instead a complementary type of charge ordering which is bond-centered, i.e. ordered on the oxygen sites. We will discuss the implications of this result for the nickelates in Sec.~\ref{sec:conclude}. 

Beyond the context of the nickelates, our model is of interest on its own as a problem in one-dimensional physics, and of the underscreened Kondo effect. Recent work has shown the possibility of gapless one-dimensional ``spin liquid'' phases, which are strikingly even more non-quasiparticle in nature than the usual Luttinger liquid. Such phases, have been found in rather exotic models with very large ring exchange interactions.\cite{DNSheng:PRB:2009} Remarkably, we find evidence for such a spin liquid-like phase in our rather simple and physically motivated Kondo lattice model (technically, we obtain a phase which appears to be proximate to the ``spin Bose metal'' phase of Ref.~\onlinecite{DNSheng:PRB:2009}, rather than this phase itself). This suggests an intriguing prospect of observing quantum spin liquid states in two and three dimensional Kondo lattice systems appropriate to real materials. Again, this idea has already been suggested in the literature,\cite{coleman1999kondo,kikoin1994mechanism,senthil2003fractionalized} but connections do not seem to have been made to realistic model Hamiltonians.

The remainder of this paper is organized as follows. In Sec.~\ref{sec:H_tJJ'} we introduce our model Hamiltonian as well as a simple semi-classical treatment that would guide our intuition about the quantum case. In Sec.~\ref{sec:DMRG} we present our DMRG results, which includes the quantum ground-state phase diagram and various correlation function in each phase. In Sec.~\ref{sec:MF} we introduce a mean-field picture, which, together with degenerate perturbation theory, explains most phases in the DMRG phase diagram. In Sec.~\ref{sec:spiral} we focus on the one phase that cannot be explained in the previous section, which we argue is a phase in proximity to the spin Bose metal phase in Ref.~\onlinecite{DNSheng:PRB:2009}. Further discussions and conclusions are presented in Sec.~\ref{sec:conclude}.

\section{The \texorpdfstring{$t$-$J$-$J'$}{t-J-J'} Hamiltonian} \label{sec:H_tJJ'}

\begin{figure}
\begin{center}
\includegraphics[scale=0.8]{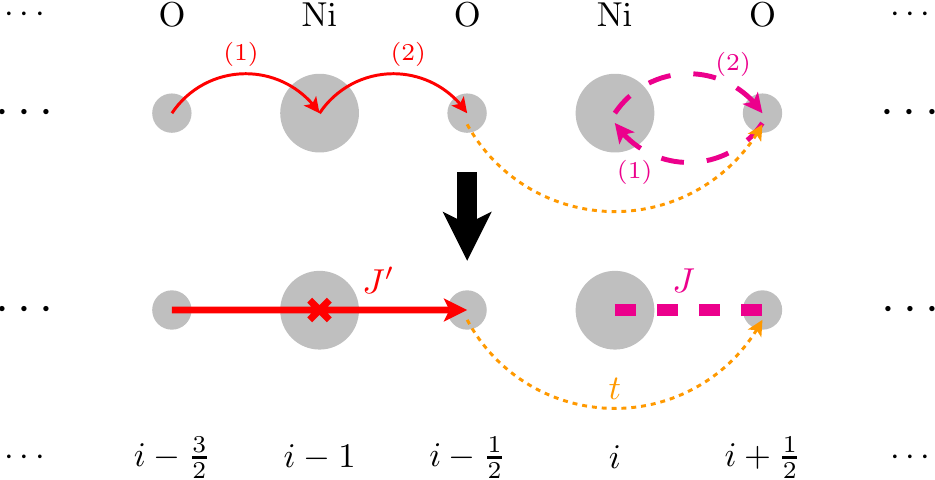}
\caption{\label{fig:H_tJJ'} (Color online) The nickelate chain and the $t$-$J$-$J'$ Hamiltonian (bottom panel), as derived from a Hubbard-typed model (top panel).}
\end{center}
\end{figure}

Motivated by the nickelate, we consider a one-dimensional (1d) chain consisting of alternating electron and spin-1 sites, as illustrated in Fig.~\ref{fig:H_tJJ'}, in which the electrons are half-filled. We shall continue to refer to the electron sites as the oxygen sites and the spin-1 sites as the nickel sites, even though our model is no longer constrained by material details. 

The interactions between the electrons and the nickel spins arise from virtual hoppings. At second order, after hopping to the nickel site, the electron or hole can either hop back to the original oxygen site or to the next oxygen site, which gives rise to two distinct contributions to the Hamiltonian. In addition, the electron can also hop directly from one oxygen site to the next without the mediation of the nickel spin in between. Taken together, this leads to the $t$-$J$-$J'$ Hamiltonian:
\begin{align} \label{eq:H_tJJ'}
H &_{tJJ'} = 
	- t \sum_i c^\dg_{i+\frac{1}{2}, \alpha} c^{\ndg}_{i-\frac{1}{2}, \alpha} + h.c. \notag \\
& + J \sum_{i} \vv{S}_i \cdot \left( 
	c^\dg_{i+\frac{1}{2}, \alpha} \frac{\vvsym{\sigma}^{\alpha\beta}}{2} c^{\ndg}_{i+\frac{1}{2}, \beta}  
	+ c^\dg_{i-\frac{1}{2}, \alpha} \frac{\vvsym{\sigma}^{\alpha\beta}}{2} c^{\ndg}_{i-\frac{1}{2}, \beta} 
	\right) \notag\\
& + J' \sum_{i} \vv{S}_i \cdot \left( 
	c^\dg_{i+\frac{1}{2}, \alpha} \frac{\vvsym{\sigma}^{\alpha\beta}}{2} c^{\ndg}_{i-\frac{1}{2}, \beta}  
	+ c^\dg_{i-\frac{1}{2}, \alpha} \frac{\vvsym{\sigma}^{\alpha\beta}}{2} c^{\ndg}_{i+\frac{1}{2}, \beta}  
	\right) \notag \\
&\phantom{_{tJJ'}} =  
	- t H_{t} + J H_{J} + J' H_{J'}\punct{,}
\end{align}
where $c^{\ndg}_{i\pm\frac{1}{2}, \alpha}$ and $c^\dg_{i\pm\frac{1}{2}, \alpha}$ are the electron operators on the oxygen site at $i\pm\frac{1}{2}$ with spin index $\alpha$, $\vv{S}_i$ is the $S=1$ spin operator on the nickel site $i$, and $\vvsym{\sigma}$ is the vector of the usual Pauli sigma matrices. $H_t$, $H_J$, and $H_{J'}$ are defined in the obvious way. For convenience, we also define $\vv{s}_{i\pm\frac{1}{2}} = c^\dg_{i\pm\frac{1}{2}\alpha} \frac{\vvsym{\sigma}^{\alpha\beta}}{2} c^{\ndg}_{i\pm\frac{1}{2}\beta}$. i.e., $\vv{s}_{i\pm\frac{1}{2}}$ is the electron spin operator on site $i\pm\frac{1}{2}$. Note that here and henceforth the spin indices are assumed to be appropriately summed. 

It is worth noting that the somewhat unfamiliar $J'$ term in $H_{tJJ'}$, is analogous to the density dependent hopping term that is customarily neglected in the $t$-$J$ model.\cite{Auerbach:Inbook:1994} However, unlike in the $t$-$J$ model, for which at low-doping this term can be approximated by a density \emph{independent} hopping, in the present case the spin variable on the nickel site remains strongly fluctuating even at low energy and thus the $J'$ term cannot be neglected. This will become evident in Sec.~\ref{sec:DMRG}.

It should also be noted that $J$ and $J'$ are in general related to each other through the parameters of the underlying Hubbard-type Hamiltonian (in which charge fluctuation is put back to the nickel sites).  We expect that $J$ and $J'$ are of the same order, but their ratio depends on microscopic details which are not known. Hence, in this paper we will treat $J$ and $J'$ as individual parameters without worrying about how they may arise from a Hubbard-type Hamiltonian. We shall, however, restrict ourselves to the quadrant in which $J, J' \geq 0$.

To gain some intuitions about the $t$-$J$-$J'$ model, consider a semi-classical treatment in which the nickel spins are treated as classical while the electrons remain quantum mechanical. For any configuration of nickel spins, $H_{tJJ'}$ reduces to a quadratic Hamiltonian of the electrons at half filling, and the ground-state configuration of the nickel spins is determined by minimizing $\avg{H_{tJJ'}}$ with respect to all possible nickel spin configurations.

For simplicity we consider only spiral configurations of the form $\vv{S}_i = S \left( \cos(q x_i) \uv{x} + \sin(q x_i) \uv{y} \right)$ for the nickel spins. In such case, the diagonalization of the quadratic Hamiltonian is facilitated by the transformation $c^{\ndg}_{i+\frac{1}{2},\alpha} \rightarrow e^{\pm i q x_i/2} c^{\ndg}_{i+\frac{1}{2},\alpha}$, where the $+$ ($-$) sign holds for $\alpha = \up$ ($\dn$). Such transformation removes the position dependence in the coefficients of the quadratic Hamiltonian obtained from $H_{tJJ'}$, which in turns can be diagonalized by a simple Fourier transform.

\begin{figure}
\begin{center}
\includegraphics[scale=0.7]{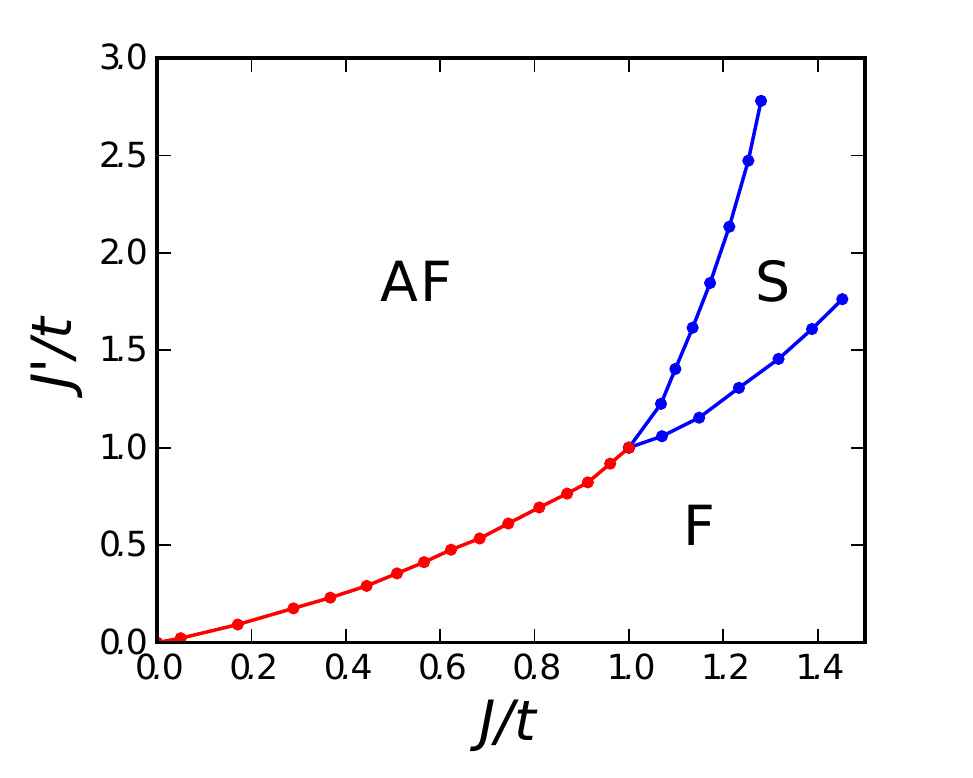}
\caption{\label{fig:semi-classical} (Color online) Phase diagram obtained from the semi-classical computation, where AF stands for a antiferromagnetic state in which wavevector $q = \pi$, F stands for a ferromagnetic phase in which wavevector $q = 0$, and S stands for a spiral phase in which $0 < q < \pi$. Note that all three phases meet at the point $J = J' = t$. }
\end{center}
\end{figure}

Carrying out the minimization with respect to $q$, we find the phase diagram shown in Fig.~\ref{fig:semi-classical}, in which the nickel spins form an antiferromagnet (AF; $q = \pi$) when $J' \gg J$, a ferromagnet (F; $q = 0$) when $J \gg J'$, and a spiral (S; $0 < q < \pi$) state when $J \approx J' \gtrsim t$.

However, it is well-known that under general circumstances the continuous symmetries of a 1d quantum system cannot be spontaneously broken. In particular, in the well-known antiferromagnetic $J_1$-$J_2$ model, the spiral phase of the classical model turns into a dimer phase when the classical spins are replaced by quantum $S = 1/2$ spins. In the dimer phase, the spin-spin correlation $\avg{\vv{s}_{-k} \cdot \vv{s}_{k}}$ exhibits no singularities but only peaks, first at $q=\pi$  and then at an incommensurate wavevector as $J_2$ increases.\cite{Bursill:JPCM:1995, White:PRB:1996} With this in mind, it is natural to ponder the fate of the spiral phase in the present model when the quantum mechanical effects are taken into account.

\section{DMRG Computations and Results} \label{sec:DMRG}

\begin{figure}
\begin{center}
\includegraphics[scale=0.8]{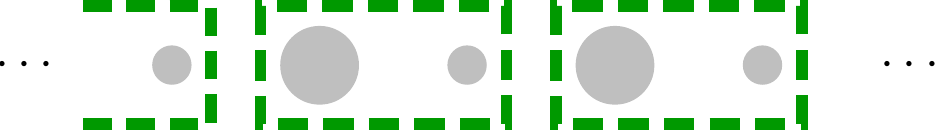}
\caption{\label{fig:unitcells} (Color online) DMRG block used in the calculations.}
\end{center}
\end{figure}

We determine the quantum ground-state phase diagram of the $t$-$J$-$J'$ Hamiltonian, Eq.~(\ref{eq:H_tJJ'}), by large-scale density-matrix renormalization group (DMRG)\cite{White:PRL:1992} calculations. We consider systems with up to $\N=96$ unit cells, each consisting of one oxygen and one nickel site (see Fig.~\ref{fig:unitcells} for illustration). For a fixed number of states kept, we found that this ``super-block'' configuration produces better convergence than the usual practice of alternating blocks of oxygen and nickel sites. In our DMRG calculation, we use open boundary conditions (OBC), and keep up to $m=4000$ states in each DMRG block. This is found to give excellent convergence in the measurements such as the ground state energy and various correlation functions with a total error of the order of or less than $10^{-6}$. The phase boundaries in the ($J/t$, $J'/t$) parameter space are determined by extensive scans of the derivatives of the ground state energy and by monitoring the correlation functions, as well as the corresponding order parameters. 

To determine the properties of the ground states, we calculate the nickel spin-spin correlation $\avg{S^{a}_k S^{a}_{-k}}$, the electron spin-spin correlation $\avg{s^{a}_k s^{a}_{-k}}$, the electron density $\avg{n_{i}}$, the electron density-density correlation $\avg{\delta n_k \delta n_{-k}}$ ($\delta n$ is defined in position space by $\delta n_{i} = n_{i} - 1$), and the nearest-neighbor nickel-oxygen spin-spin correlation $\avg{\vv{s}_{i\pm\frac{1}{2}} \cdot \vv{S}_i}$. To determine whether the phase is conducting, we also calculate the charge gap $\Delta=[E_0(N+2)+E_0(N-2)-2E_0(N)]/2$. Finally, we also obtain the central charge $c$ by calculating the von Neumann entanglement entropy $S_A$ for a partition of the system into two halves of length $\ell$, $L-\ell$, with varying length $\ell$ of the  sub system $A$.  A universal logarithmic dependence of $S_A$ on $\ell$ is expected in the thermodynamic limit with a prefactor depending on $c$, and accurate results can usually be obtained using finite-size scaling formulas derived from conformal field theory, see Refs.\onlinecite{Calabrese:JSM:2004},\onlinecite{Calabrese:JPA:2009}. The calculation of the von Neumann entropy is computationally expensive and generally the most difficult of all physical quantities to converge, especially in highly entangled states.  Therefore our results for $c$ are more limited than for the other quantities discussed above.

\begin{figure}
\begin{center}
\includegraphics[scale=0.28]{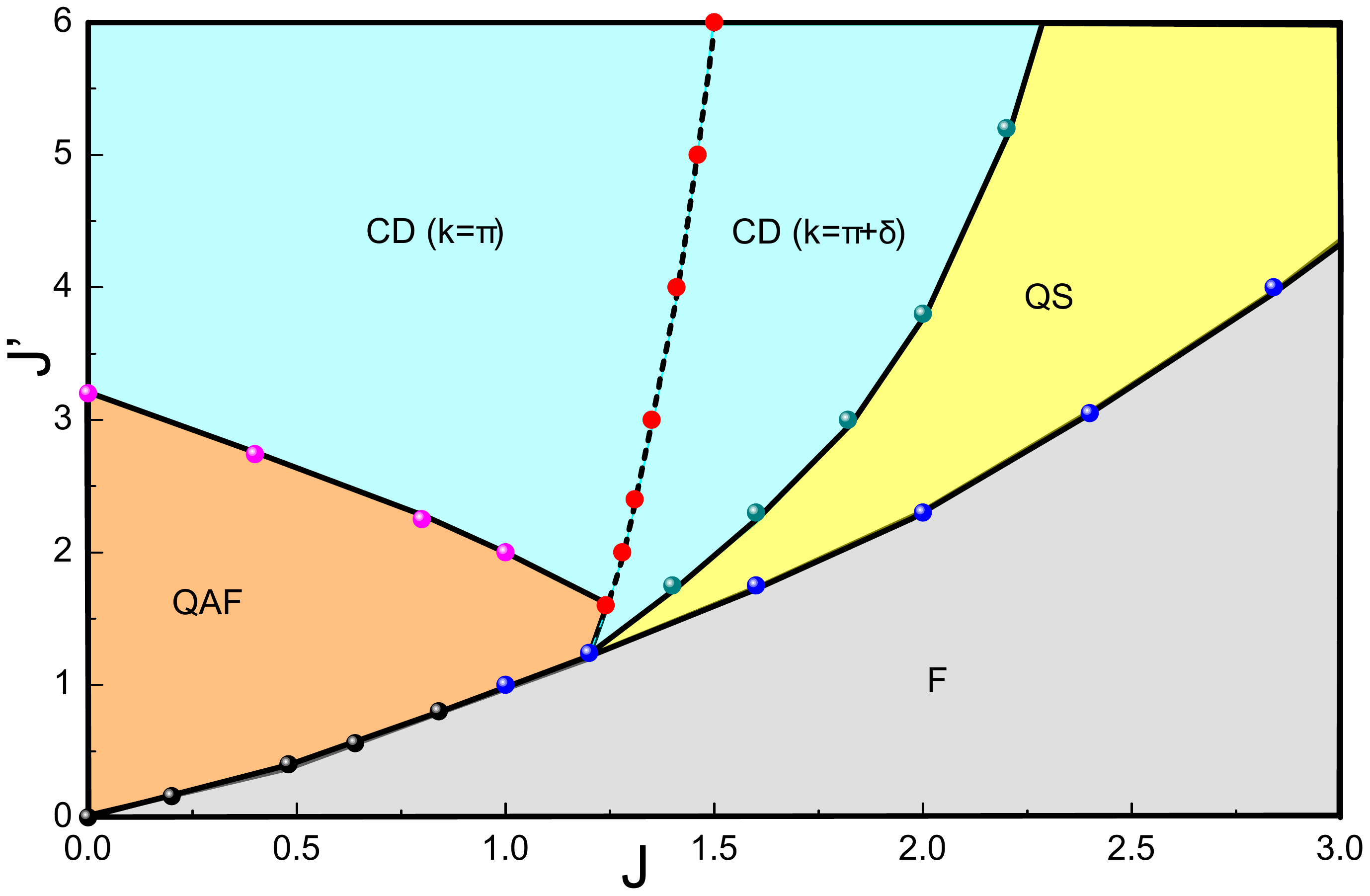}
\caption{\label{fig:DMRG_phases} (Color online) Phase diagram from the DMRG calculation. Here F stands for the ferromagnetic phase, QAF stands for the quasi-long-range antiferromagnetic phase, CD stands for the charge-density ordered phase, and QS stands for the quasi-long-range spiral phase. The dashed line within the CD phase represents the boundary in which the peak in $\avg{S^{z}_k S^{z}_{-k}}$ changes from $k = \pi$ to $k = \pi \pm \delta$.} 
\end{center}
\end{figure}

\begin{figure}
\begin{center}
\includegraphics[scale=0.28]{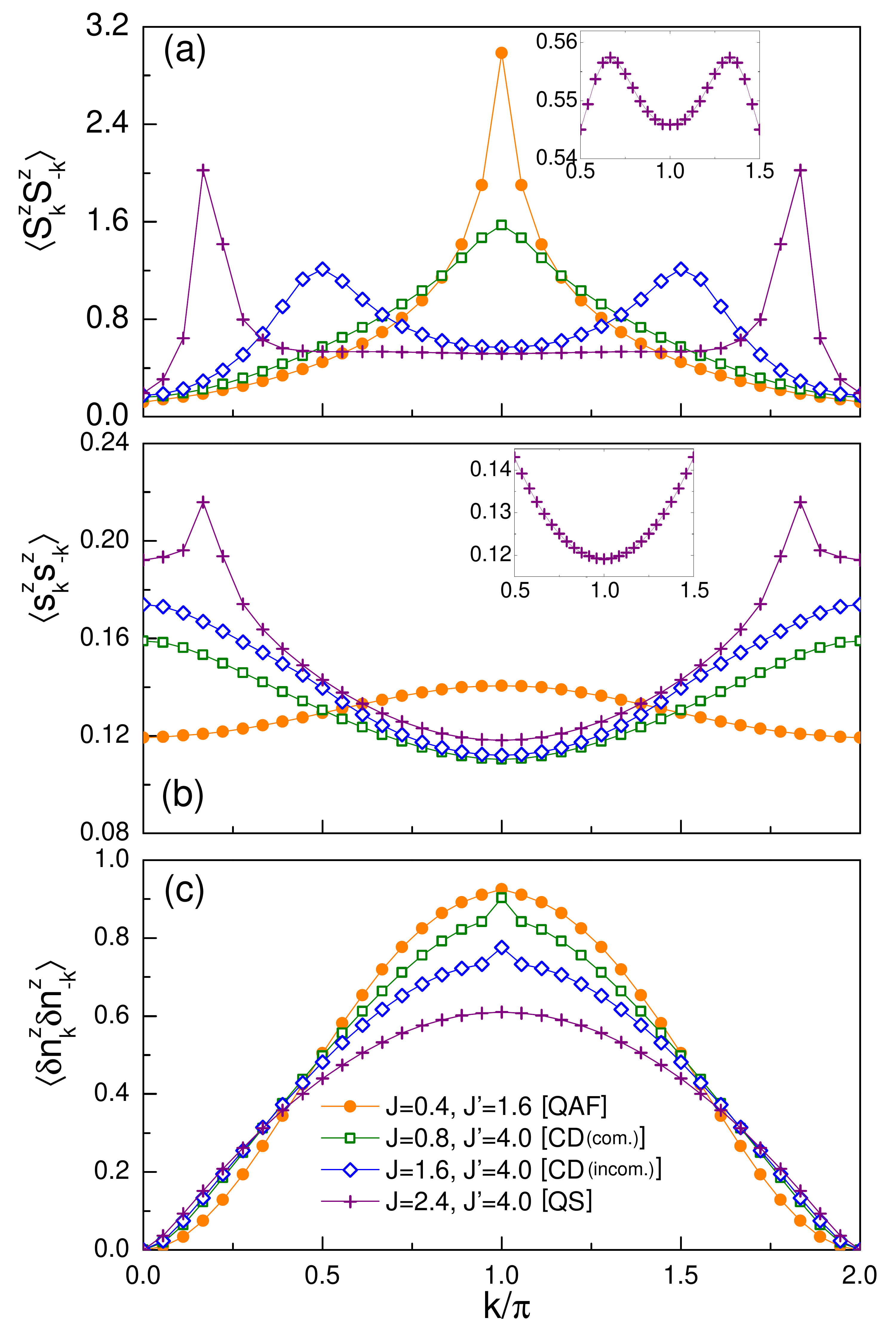}
\caption{\label{fig:correlations} (Color online) (a) Nickel spin-spin correlation $\avg{S^{z}_k S^{z}_{-k}}$, (b) electron spin-spin correlation $\avg{s^z_k s^z_{-k}}$, and (c) electron density-density correlation $\avg{\delta n_k \delta n_{-k}}$ for (orange, filled circular symbols) $J/t=0.4$ and $J'/t=1.6$, (green, empty square symbols) $J/t = 0.8$ and $J'/t = 4.0$, (blue, empty diamond symbols) $J/t = 1.6$ and $J'/t = 4.0$, and (purple, Greek cross symbols) $J/t=2.4$ and $J'/t=4.0$. The insets of panels~(a) and (b) show the details of the respective correlation for $J/t=2.4$ and $J'/t=4.0$ in the range $0.5 < k/\pi < 1.5$. }
\end{center}
\end{figure}

\begin{figure}
\begin{center}
\includegraphics[scale=0.28]{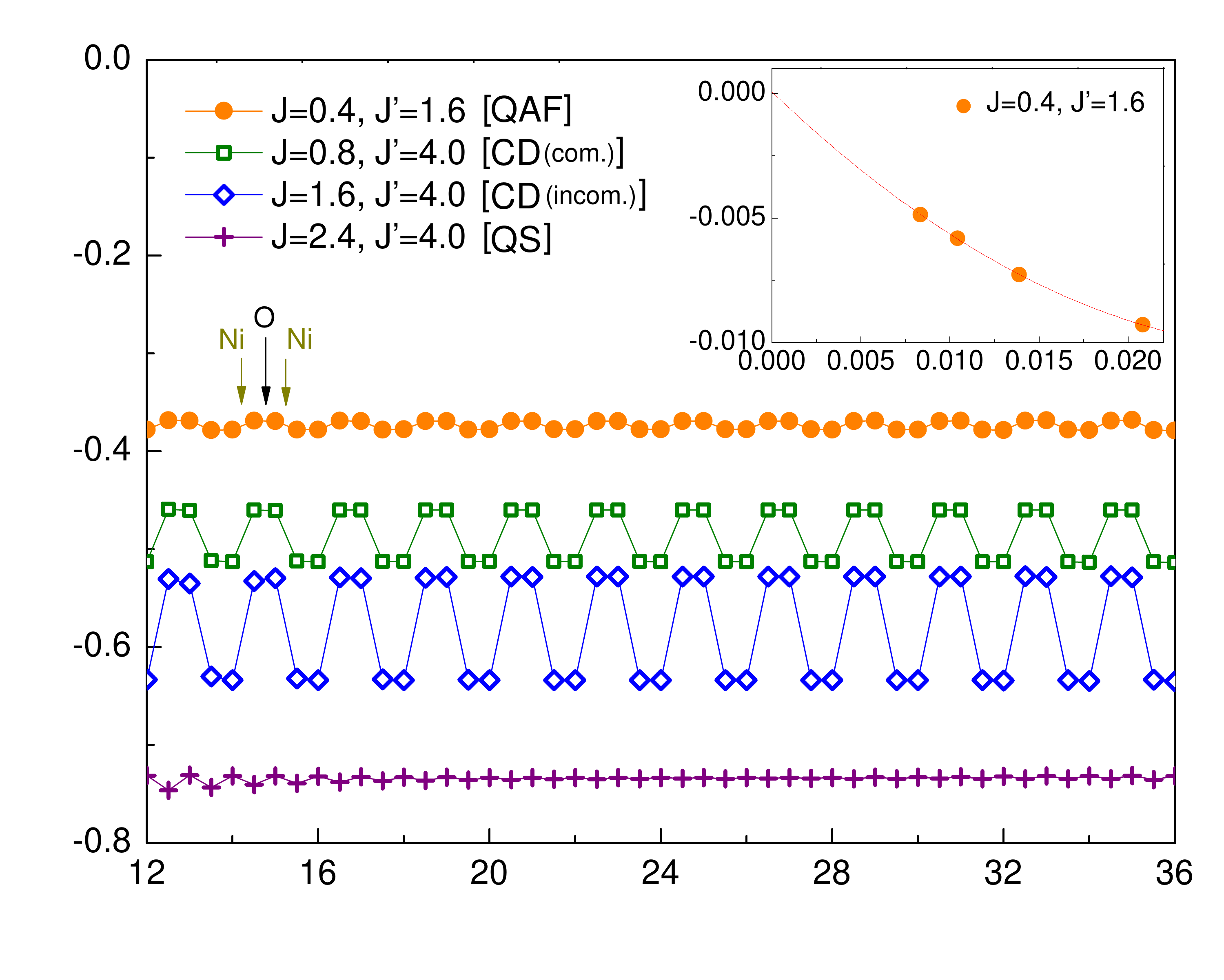}
\caption{\label{fig:Ss_DMRG} (Color online) The nearest-neighbor nickel-oxygen spin-spin correlation $\avg{\vv{s}_{i\pm\frac{1}{2}} \cdot \vv{S}_i}$ for the characteristic points in parameter space used in Table~\ref{tbl:D_and_c}, with system size $N = 48$. The two correlations are combined and the index $i$ now labels the bond center of the correlation. The positions of the nickel and the oxygen sites are indicated by the labels above the curves. The inset shows finite size scaling of the oscillation amplitude in $\avg{\vv{s}_{i\pm\frac{1}{2}} \cdot \vv{S}_i}$ for $J = 0.4$ and $J' = 1.6$ using a second order polynomial fit. }
\end{center}
\end{figure}

Our main result is the phase diagram presented in
Fig.~\ref{fig:DMRG_phases}, in which we find four distinct phases as $J$
and $J'$ are varied. In addition, the correlation functions at characteristic points in parameter space are shown in Fig.~\ref{fig:correlations} and \ref{fig:Ss_DMRG}. Note that we have only shown the $zz$-components of the spin-spin correlation functions. We remark that the phases labeled as QAF, CD, and QS (detailed below) possess spin $SU(2)$ invariance, and thus the other components of the spin-spin correlations are essentially identical to the ones we showed. For the ferromagnetic (F) phase, the $z$-axis in spin space is constrained to $M_z = 0$, and thus the magnetization axis is perpendicular to the $z$-axis.

\begin{table}
\caption{\label{tbl:D_and_c} Charge gap $\Delta$ and central charge $c$ for characteristic parameter values within each phase of the phase diagram Fig.~\ref{fig:DMRG_phases}. For the CD phase, ``com.'' indicates peak in $\avg{S^{z}_k S^{z}_{-k}}$ at $k = \pi$ while ``incom.'' indicates peak in $\avg{S^{z}_k S^{z}_{-k}}$ at incommensurate wavevectors. }
\begin{ruledtabular}
\begin{tabular}{cccccc}
$J/t$ & $0.4$ & $0.8$ & $1.6$ & $2.4$ & $2.0$ \\
$J'/t$ & $1.6$ & $4.0$ & $4.0$ & $4.0$ & $0.4$ \\
\hline
Phase & QAF & CD (com.) & CD (incom.) & QS & F\\
$\Delta/t$ & $1.14$ & $0.96$ & $0.68$ & 0.21 & $1.96$\\
 $c$ & $0.8$ & $0$ & $0$ & \textbf{Unknown} & 0
\end{tabular}
\end{ruledtabular}
\end{table}

\begin{figure}
\begin{center}
\includegraphics[scale=0.28]{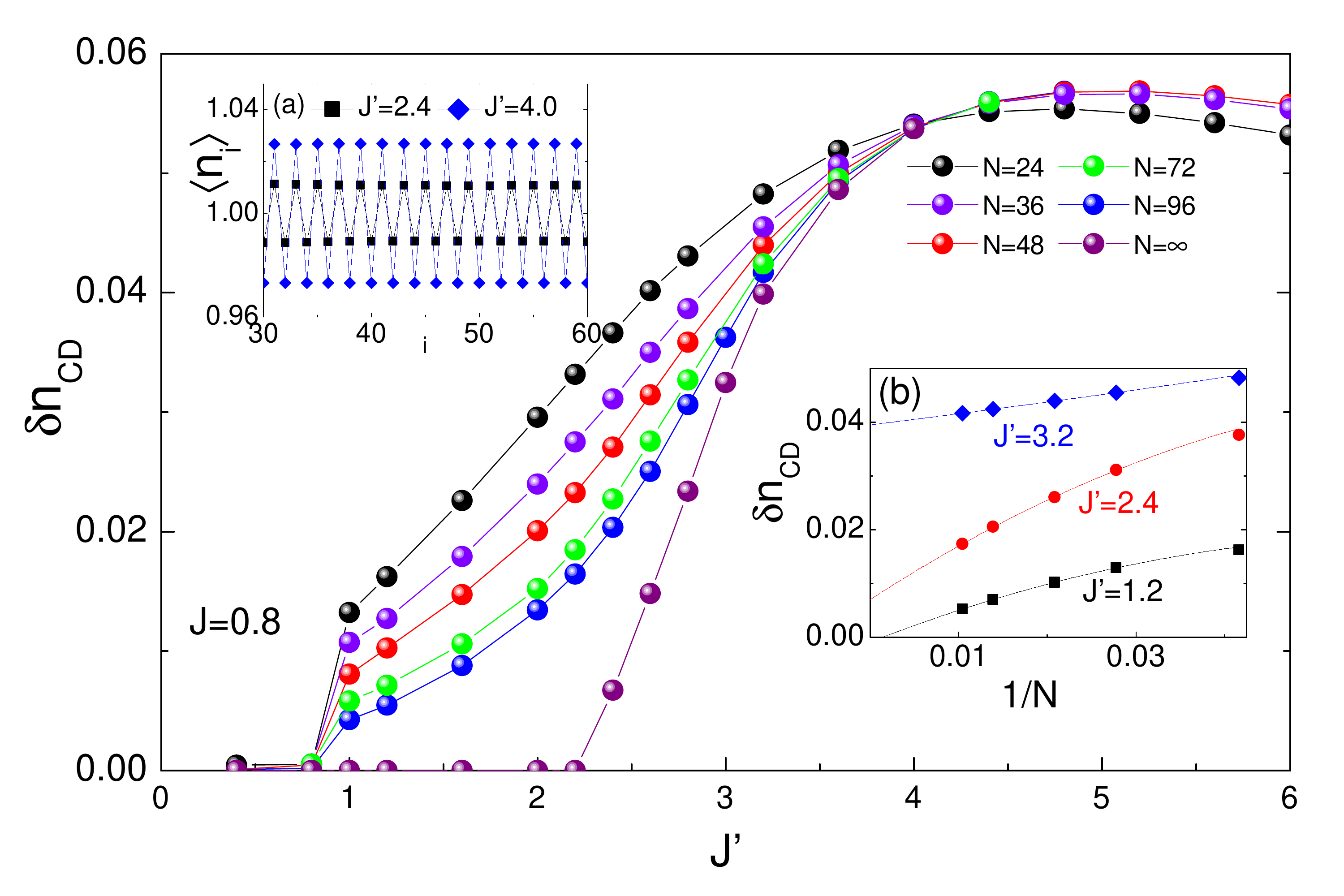}
\caption{\label{fig:CDW} (Color online) The development of charge-density wave order parameter  $\delta n_{\txt{CD}}=\avgnormal{n_{\frac{\N}{2}}} - \avgnormal{n_{\frac{\N}{2}+1}}$ as $J'/t$ increases while $J/t =0.8$ is fixed. The inset (a) shows $\avg{n_i}$ as function of unit-cell index $i$ for (black, filled square symbols) $J/t=0.8$ and $J'/t = 2.4$ and (blue, filled diamond symbols) $J/t=0.8$ and $J'/t = 4.0$, with system size $N = 96$. The inset (b) shows examples of finite-size scaling of $\delta n_{\txt{CD}}$ at different $J'$ using second-order polynomials.} 
\end{center}
\end{figure}

For $J \gtrsim J'$ the system exhibits ferromagnetic order with magnetization $M \approx 0.5~ \mu_B$ per unit cell.  This is consistent with previous works in the literature on the alternating spin-1--spin-1/2 chain,\cite{Brehmer:JPCM:1997, Pati:PRB:1997} which can be thought of as the $J'=0, J\rightarrow\infty$ limit of the present model.  For $J$ and $J'$ both small and $J' \gtrsim J$, the system exhibits quasi-long-range antiferromagnetic (QAF) order characterized by a sharp peak in $\avg{S^{z}_k S^{z}_{-k}}$ at $k = \pi$, while $\avg{s^{z}_k s^{z}_{-k}}$ and $\avg{\delta n_k \delta n_{-k}}$ remain largely featureless. As $J'$ further increases such that $J' \gtrsim J \approx t$, the system develops a charge-density (CD) order with period 2 as shown in inset (a) in Fig.~\ref{fig:CDW}, and the sharp singularity in $\avg{S^{z}_k S^{z}_{-k}}$ in the QAF phase becomes a broad peak. As $J$ increases within this CD phase, the peak originally located at $k = \pi$ splits into two peaks at $k = \pi \pm \delta$, with $\delta$ increasing as $J$ increases.

\begin{figure}
\begin{center}
\includegraphics[scale=0.28]{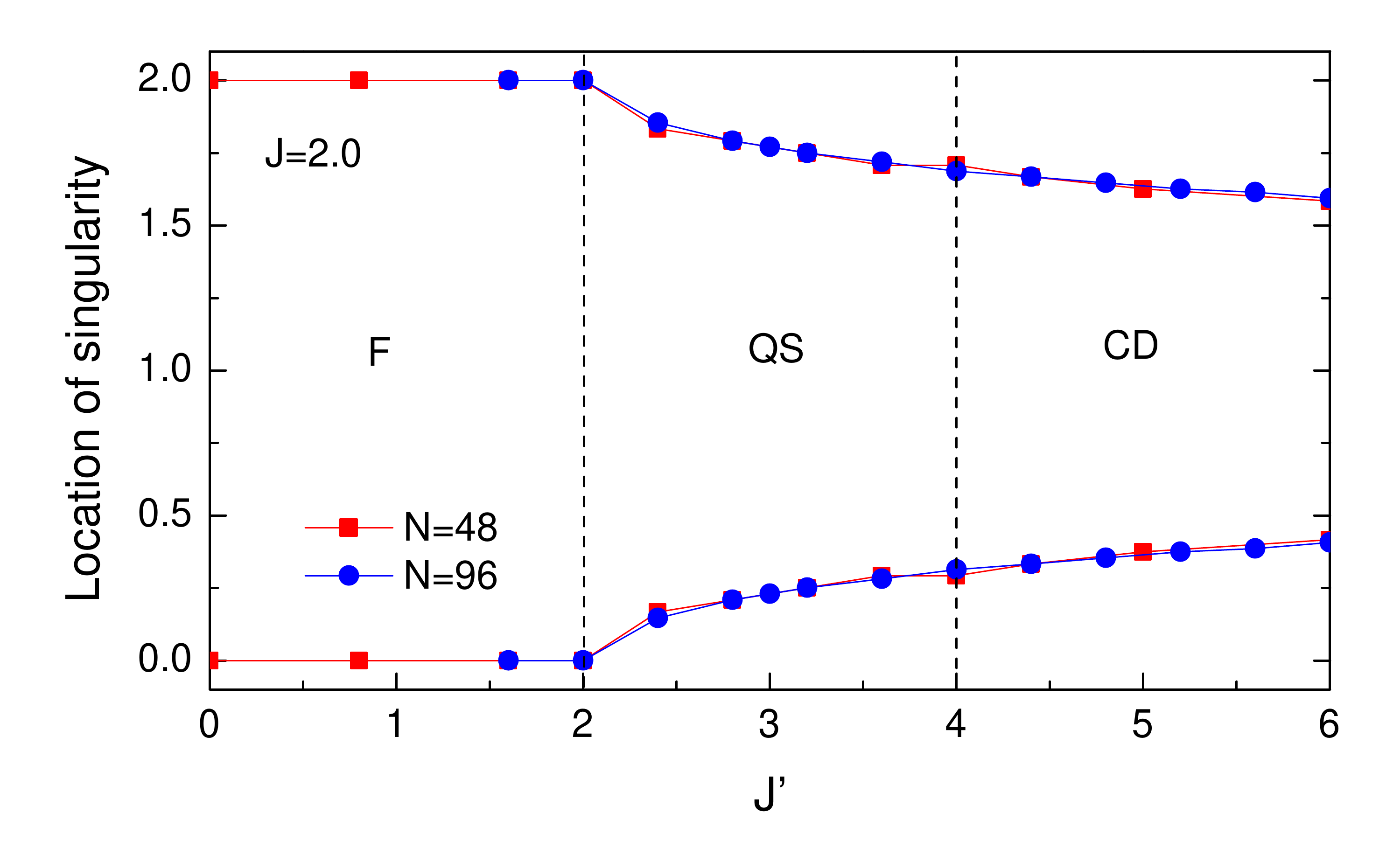}
\caption{\label{fig:peak_location} (Color online) Locations of the singularity (resp. peak) in the ferromagnetic and QS phases (resp. CD phase) as function of $J'/t$ when $J/t=2.0$ is fixed.}
\end{center}
\end{figure}

Most interestingly, when $J$ and $J'$ are both large and $J' \gtrsim J$ the system seems to exhibit quasi-long-range spiral (QS) order characterized by sharp singularities in \emph{both the nickel spin-spin correlation $\avg{S^{z}_k S^{z}_{-k}}$ and the \emph{electron} spin-spin correlations $\avg{s^{z}_k s^{z}_{-k}}$ at incommensurate vectors $k = \pi \pm \delta$}, while the electron density correlation $\avg{\delta n_k \delta n_{-k}}$ remains largely featureless, and no signs of CD order are found. Moreover, when the correlation functions are examined more carefully near $k=\pi$, a tiny broad peak can be discerned at wavevector roughly equal to $\pm \delta$ in $\avg{S^{z}_k S^{z}_{-k}}$, while no such features are found in $\avg{s^{z}_k s^{z}_{-k}}$. It should be remarked that the value of $\delta$ evolves continuously as $J'$ and $J$ changes and connects across the phase boundaries into the ferromagnetic and the CD phases, as shown in Fig.~\ref{fig:peak_location}. 

In addition, from Fig.~\ref{fig:Ss_DMRG} we see that in the QAF and the QS phases, the nearest-neighbor spin-spin correlation $\avg{\vv{s}_{i\pm\frac{1}{2}} \cdot \vv{S}_i}$ is essentially uniform in the thermodynamic limit if not already in finite-size systems, while in the CD phase $\avg{\vv{s}_{i\pm\frac{1}{2}} \cdot \vv{S}_i}$ has a significant two-unit-cell oscillation centered at the oxygen site, which is consistent with the symmetry of the CD order. Note that the oscillations of $\avg{\vv{s}_{i\pm\frac{1}{2}} \cdot \vv{S}_i}$ in the CD phase have larger amplitudes than the corresponding oscillations in $\avg{n_i}$, and that in general $\left|\avg{\vv{s}_{i\pm\frac{1}{2}} \cdot \vv{S}_i}\right|$ increases as the parameters $(J/t, J'/t)$ increase.

\begin{figure}
\begin{center}
\includegraphics[scale=0.28]{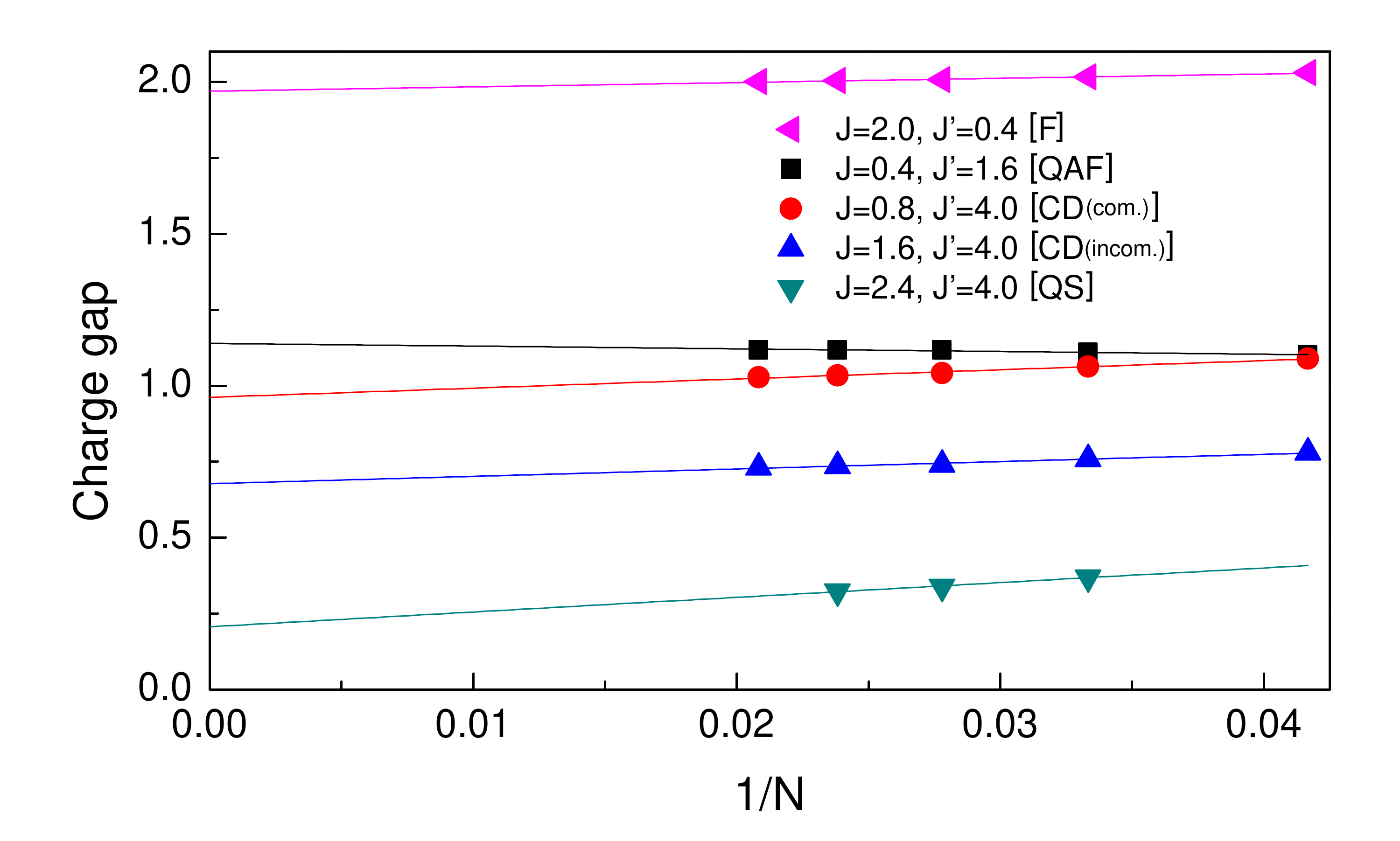}
\caption{\label{fig:finite_size_D} (Color online) Finite size scaling of the charge gap for the characteristic points in parameter space used in Table~\ref{tbl:D_and_c}.}
\end{center}
\end{figure}

\begin{figure}
\begin{center}
\includegraphics[scale=0.28]{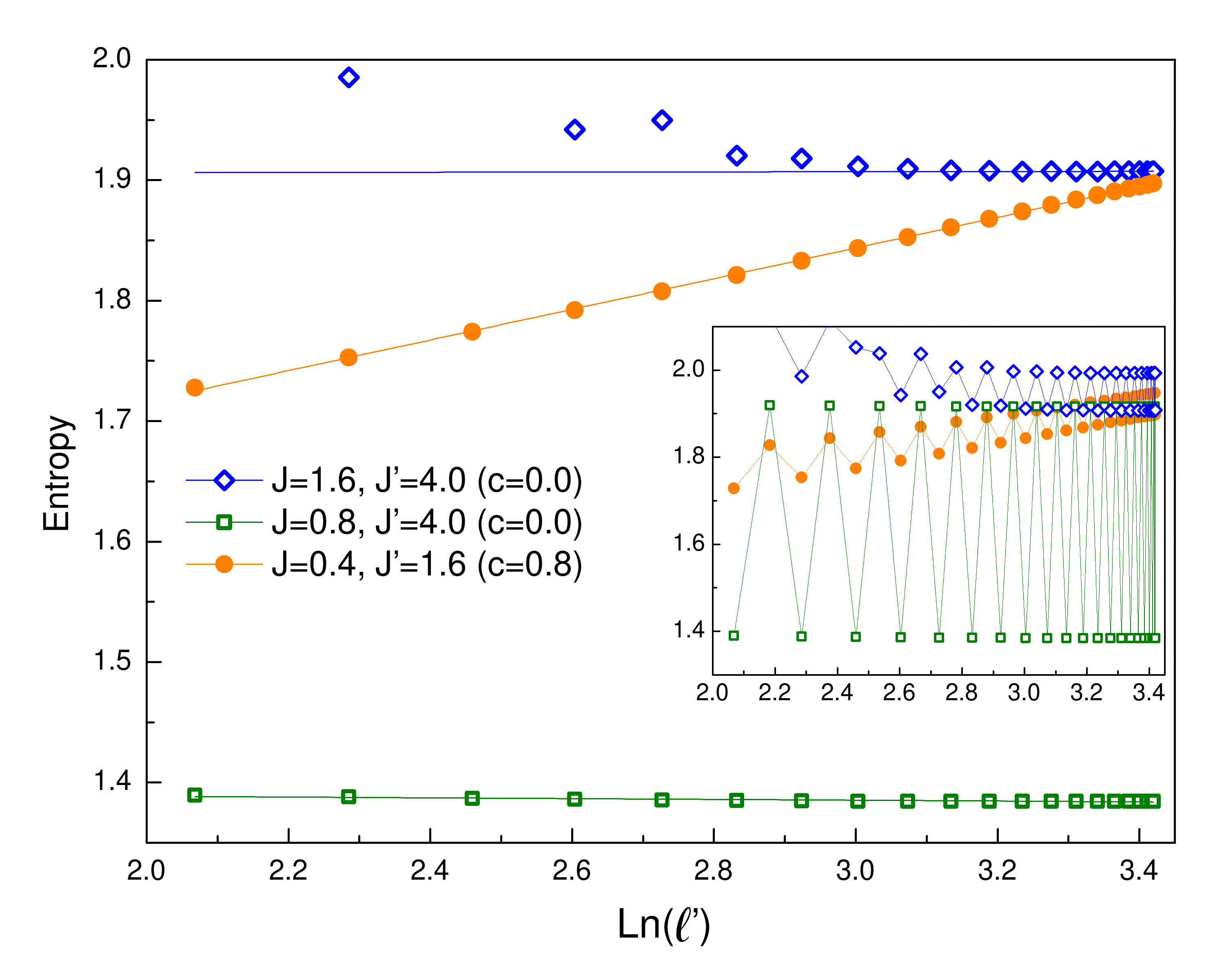}
\caption{\label{fig:EE_scaling} (Color online) The lower branch of the von
  Neumann entanglement entropies, obtained from the lower envelope of
  the full entropy shown in the inset, as a function of the
  conformally-transformed length\cite{Calabrese:JPA:2009} $\ell^\prime = (L/\pi)\sin (\pi \ell/L)$ of the subsystem $A$ for the
  characteristic points in parameter space in
  Table~\ref{tbl:D_and_c} ($L$ is the system size). The solid lines are linear fits using
  $S_A=\frac{c}{6}\ln \ell^\prime$, which determines the central
  charge $c$. } 
\end{center}
\end{figure}

To shed further light into the phases we obtained in DMRG, we show the charge gap $\Delta$ and central charge $c$ for characteristic parameter values within each phase in Table~\ref{tbl:D_and_c} (computational details are illustrated in Figs.~\ref{fig:finite_size_D} and \ref{fig:EE_scaling}). From the table it can be seen that all phases have finite charge gaps. Moreover, the values of the central charges suggest that one bosonic degree of freedom remains gapless in the QAF phase, while no degree of freedom remains gapless in the CD and F phase. Unfortunately, owing to the large (twelve) internal dimension of the unit cell, we cannot accurately determine the central charge of the QS phase. \footnote{We are unable to converge the entanglement entropy in the QS phase.  This is signaled by that fact that it continues to grow with the number of states kept in the DMRG, even when this number reaches the maximum possible given the technical limitations of our simulation.  Moreover, there are significant incommensurate oscillations in the entanglement entropy in the QS phase, and we need data on systems much larger than the quasi period of these oscillations to get reliable results.   This period is long, which makes the calculations even more challenging.}

\section{Mean-field Kondo picture and effective spin-1/2 model} \label{sec:MF} 

In this section we introduce a slave-particle representation of the nickel spin, from which we obtain a quadratic mean-field theory (Sec.~\ref{subsec:slave}). Assuming the simplest form of mean-field ansatz, this leads to a spectrum that contains a flat band exactly at the Fermi energy. We then show that such a degeneracy can be lifted using second-order degenerate perturbation theory, which allows us to map the present problem to an effective spin-1/2 model (Sec.~\ref{subsec:perturb}). We then carry out an explicit calculation of the parameters in the effective spin-1/2 model, and compare a quantum mean-field analysis of the phase diagram of this model with the DMRG results (Sec.~\ref{subsec:J_ij}). Our main result in this section is the quantum mean-field phase diagram of Fig.~\ref{fig:J1J2_grid}.

\subsection{Slave-particle representation and its mean field} \label{subsec:slave}

To interpret the DMRG results, it is useful to introduce a mean-field picture. Specifically, we want mean-field ``parent states'' that are translationally and spin $SU(2)$ invariant, under which the quasi-long-range orders and/or the broken symmetries manifest as singularities in the correlation functions in the resulting low-energy effective theory. To this end, we consider the following Schwinger fermion representation, in which the spin-1 operators $\vv{S}_i$ are written in terms of four species of spinons $f^{\ndg}_{ia\alpha}$, carrying both orbital indices (indicated by lowercase Latin characters) and spin indices (indicated by lowercase Greek characters):

\begin{equation} \label{eq:Schwinger}
\vv{S}_i = \sum_{a=1,2} f^{\dg}_{i a \alpha} \frac{\vvsym{\sigma}^{\alpha \beta}}{2} f^{\ndg}_{i a \beta} \punct{,}
\end{equation}
in which the spinons are subjected to the constraints
\begin{equation} \label{eq:constraints}
\sum_{a}  f^{\dg}_{i a \alpha} f^{\ndg}_{i a \alpha} = 2 \punct{,}\quad 
\sum_{a,b} f^{\dg}_{i a \alpha} \vvsym{\tau}^{ab} f^{\ndg}_{i b \alpha} = 0 \punct{,}
\end{equation}
(recall that the spin indices are implicitly summed), in which $\vvsym{\tau}$ is the vector of the Pauli sigma matrices that act on the orbital indices.\cite{CKXu:PRL:2012} For brevity, henceforth we shall assume that the orbital indices are appropriately summed.

Substituting the Schwinger fermion representation into the $t$-$J$-$J'$ Hamiltonian results in four-fermion terms that are schematically of the form $c^\dg f^\dg f^{\ndg}\! c^{\ndg}\!\!$, which can be handled by a Hartree-Fock-type decomposition. Together, this leads us to the following mean-field Hamiltonian:

\begin{align} \label{eq:H_MF}
H&^{\txt{MF}} = -\mu_e \sum_i c^\dg_{i+\frac{1}{2}, \alpha} c^{\ndg}_{i+\frac{1}{2}, \alpha} - t \sum_i c^\dg_{i+\frac{1}{2}, \alpha} c^{\ndg}_{i-\frac{1}{2}, \alpha} + h.c. \notag \\
& + \sum_{i} Q_a c^\dg_{i+\frac{1}{2}, \alpha} f^{\ndg}_{i a \alpha} + Q'_a c^\dg_{i-\frac{1}{2}, \alpha} f^{\ndg}_{i a \alpha} + h.c. \notag \\
& + \sum_{i}  \lambda f^{\dg}_{i a \alpha} f^{\ndg}_{i a \alpha} 
+ \vv{N} \cdot f^{\dg}_{i a \alpha} \vvsym{\tau}^{ab} f^{\ndg}_{i b \alpha}\punct{,}
\end{align} 
in which the Lagrange multipliers $\lambda$ and $\vv{N}$ are introduced to enforce the constraints Eq.~(\ref{eq:constraints}) on average, and the orbital-dependent Kondo hoppings $Q_a$ and $Q'_a$ are to be determined by the minimization of the mean-field energy; i.e., $Q_a$ and $Q'_a$ are to be chosen such that $\avg{H_{tJJ'}}_{\txt{MF}}$ is minimized (here $\avg{\cdot}_{\txt{MF}}$ denotes expectation value with respect to the mean-field state obtained from $H^{\txt{MF}}$).

Notice that $\lambda$, $\vv{N}$, $Q_a$, and $Q'_a$ carry no site indices since translation invariance is assumed. Similarly, $Q_a$ and $Q'_a$ carry no spin indices since spin $SU(2)$ invariance is assumed. By the same token, we also neglect the ``magnetic'' mean-field terms $(f^{\dg}_{i a \alpha} \vvsym{\sigma}^{\alpha\beta} f^{\ndg}_{i a \beta})$ and $(c^{\dg}_{i\pm\frac{1}{2},\alpha} \vvsym{\sigma}^{\alpha\beta} c^{\ndg}_{i\mp\frac{1}{2},\beta})$ in $H^{\txt{MF}}$.

It should be remarked that $H^{\txt{MF}}$ can alternatively be derived in a Feynman path integral approach in which fluctuating $\lambda_i$ and $\vv{N}_i$ are introduced as \emph{auxiliary fields} to the partition function, such that the constraints are enforced exactly upon functional integration. Similarly, in this approach $Q_{ia}$ and $Q'_{ia}$ are introduced as \emph{fluctuating} bosonic Hubbard-Stratonovich fields that upon functional integration reproduce the appropriate four-fermion terms.\cite{Read:JPC:1983, Read:PRB:1984, Auerbach:PRL:1986} In this context, $\lambda_i$ and $\vv{N}_i$ can also be interpreted as the temporal components of a $U(1)$ and an $SU(2)$ gauge field, respectively. Together, they corresponds to the $U(2) \sim U(1) \times SU(2)$ gauge redundancy $f_{ia\alpha} \rightarrow U_{i}^{ab} f_{ib\alpha}$ of the spinons. However, at this stage we shall take the simpler picture and treat the variables $\lambda$, $\vv{N}$, $Q_{a}$, and $Q'_{a}$ as parameters of $H^{\txt{MF}}$.

\subsection{Mean-field ans\"{a}tze, flat bands, and degenerate perturbation theory} \label{subsec:perturb}

While in general $Q_{a}$ and $Q'_{a}$ are distinct, it is natural to first consider scenarios in which $Q_a = Q'_a$, where we can make use of the $U(2)$ gauge redundancy to set $Q_{a} = Q'_{a} = r [1,0]^T$, with $r \geq 0$. Then, the constraints Eq.~(\ref{eq:constraints}) requires $\vv{N} \propto \uv{z}$ and thus the last line of Eq.~(\ref{eq:H_MF}) can be rewritten as $\sum_{ia} \lambda_a  f^\dg_{ia\alpha} f^{\ndg}_{ia\alpha}$, in which $\lambda_a = \lambda + (-1)^{a+1} N_z$. 

\begin{figure}
\begin{center}
\includegraphics[scale=0.6]{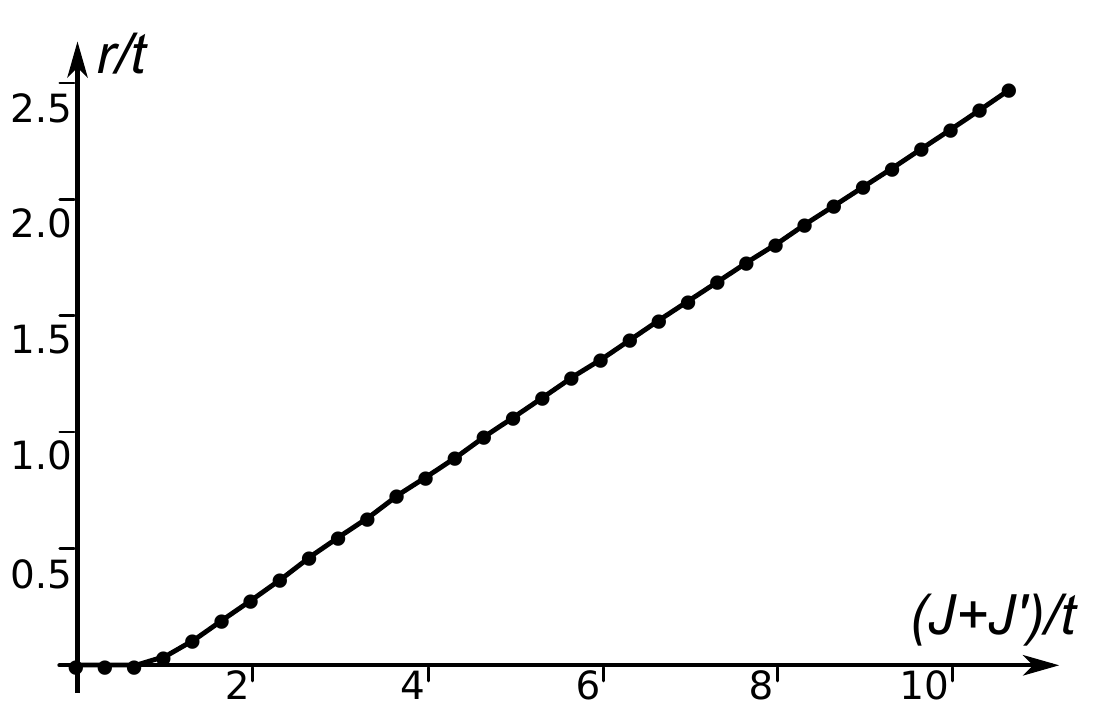}
\caption{\label{fig:r_vs_J} Dependence of $r/t$ on $(J+J')/t$ in the class of mean-field ans\"{a}tze defined by $Q_{a} = Q'_{a} = r [1,0]^T$.}
\end{center}
\end{figure}

\begin{figure}
\begin{center}
\subfigure[\label{fig:J=J'=2} $J = J' = \frac{4t}{3}$]{\includegraphics[scale=0.35]{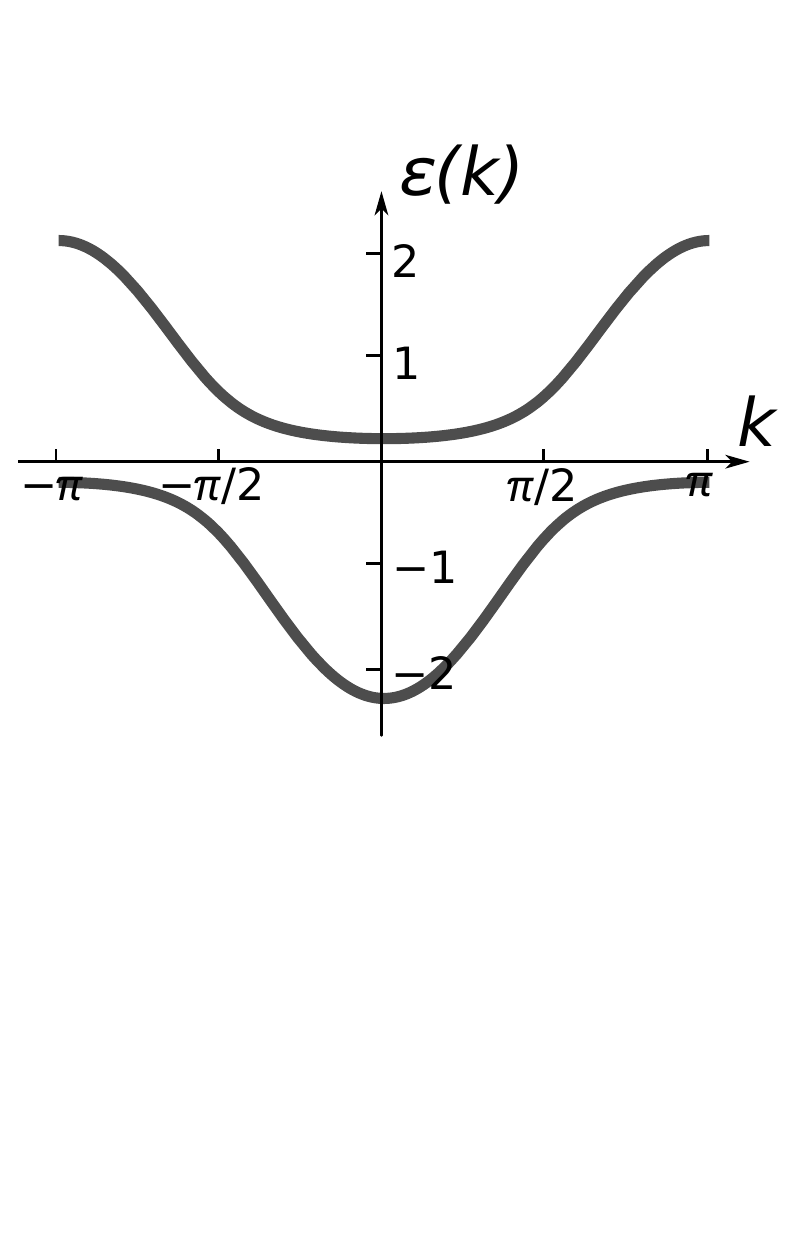}}
\subfigure[\label{fig:J=J'=4} $J = J' = \frac{8t}{3}$]{\includegraphics[scale=0.35]{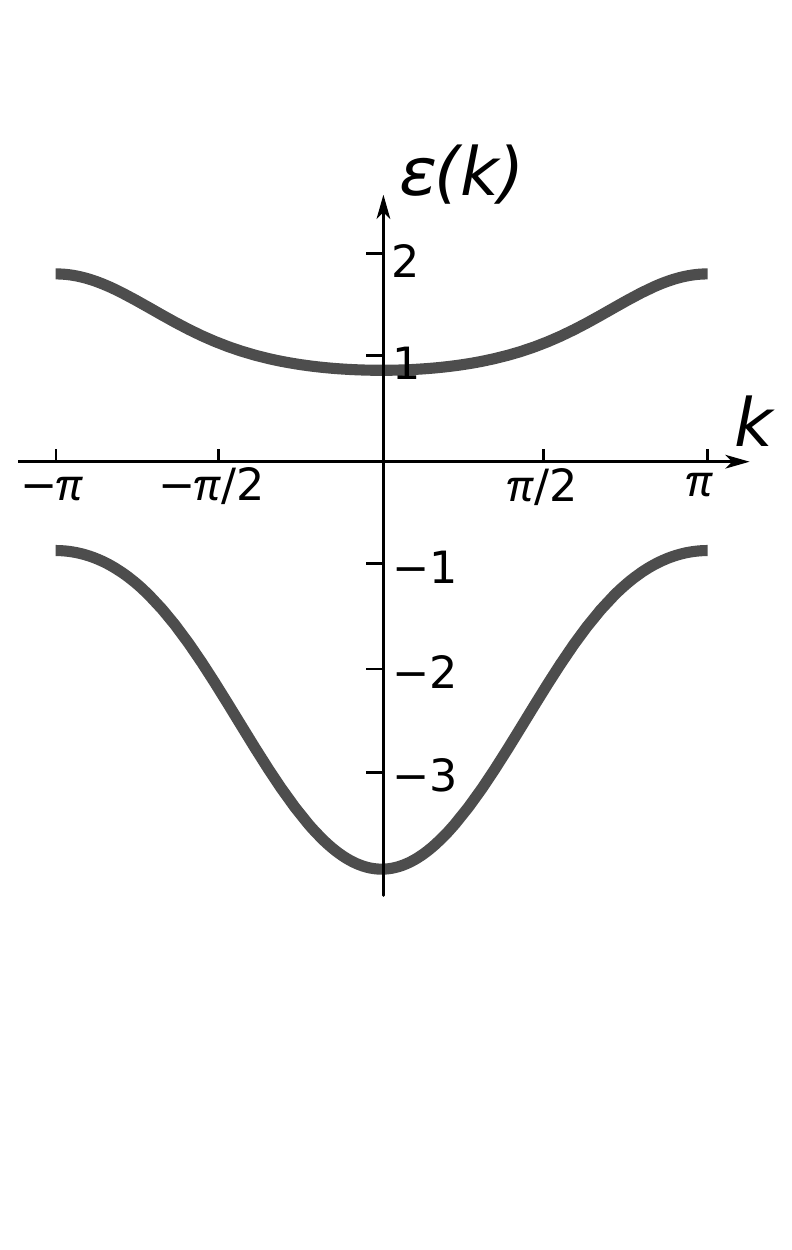}}
\subfigure[\label{fig:J=J'=8} $J = J' = \frac{16t}{3}$]{\includegraphics[scale=0.35]{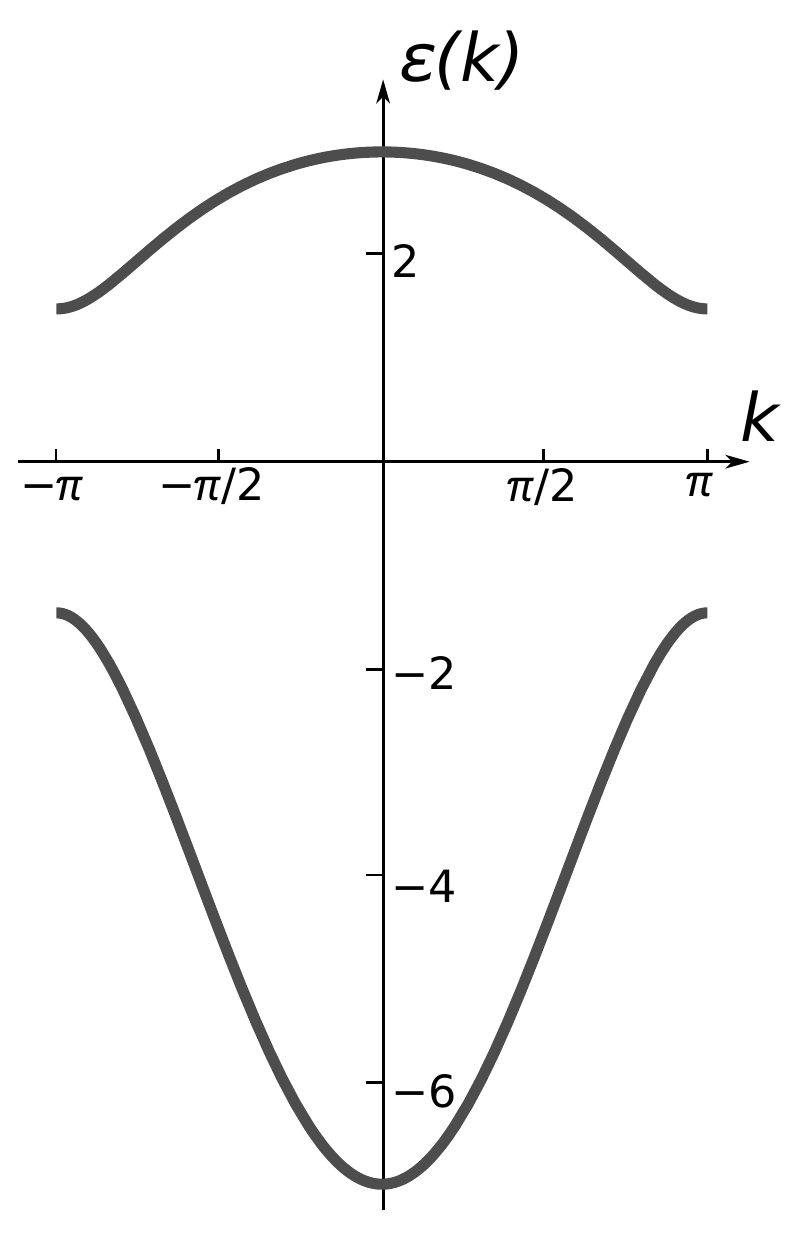}}
\caption{\label{fig:MF_spectra} Typical mean-field spectra obtained from $H^{\txt{MF}}$ for the class of mean-field ans\"{a}tze defined by $Q_{a} = Q'_{a} = r [1,0]^T$. Note that the degenerate flat bands at zero energy are not plotted. }
\end{center}
\end{figure}

From this, it can be seen that for this class of ans\"{a}tze, the $f_2$ spinons are completely decoupled from the other fermions in the system and enter $H^{\txt{MF}}$ only through a (species-dependent) chemical potential. Consequently, the $f_2$ spinons form a flat band in the mean-field spectrum, which has to be half-filled in order to satisfy the constraints. The two remaining species of fermions form a Kondo band insulator, in which the band gap is controlled by the ratio $r/t$ that increases as $J,J'$ increases. Moreover, it is easily checked that for this class of ans\"{a}tze $\avg{H_{J}} = \avg{H_{J'}}$, and hence the mean-field spectrum depends only on $(J+J')/t$. Carrying out the minimization of $\avg{H_{tJJ'}}_{\txt{MF}}$ with respect to $r$, we obtain $r/t$ as a function of $(J+J')/t$ as shown in Fig.~\ref{fig:r_vs_J}. The mean-field spectra for ans\"{a}tze with different $r/t$ are plotted in Fig.~\ref{fig:MF_spectra}.  Note that in the mean-field picture $\left|\avg{\vv{s}_{i\pm\frac{1}{2}} \cdot \vv{S}_i}\right| \propto \sum_{a} \left|\avg{f^\dg_{ia\alpha} c^{\ndg}_{i\pm\frac{1}{2}\alpha}} \right|^2 \sim r^2$. Hence, the trend of increasing $r/t$ as $(J+J')/t$ increases is consistent with the trend in DMRG (c.f.\@ Fig.~\ref{fig:Ss_DMRG}).

\begin{figure}
\begin{center}
\includegraphics[scale=0.8]{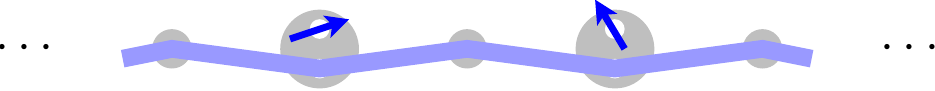}
\caption{\label{fig:Kondo_screen} (Color online) Schematic of the Kondo under screening, in which the electrons screen out only half of the nickel spin on each nickel site, leaving behind an unscreened spin-1/2.}
\end{center}
\end{figure}

Physically, the half-filled flat band can be interpreted as free $S=1/2$ spins on the nickel sites that arise from Kondo under screening, in which the electrons screen out only half of the nickel spin on each site (see Fig.~\ref{fig:Kondo_screen} for illustration). This degeneracy among the $f_2$ spinons is expected to be lifted when effects beyond mean field are considered. To capture such effects, we perform second-order degenerate perturbation theory on the mean-field state, in which the perturbation is provided by the residual interaction $H_{\txt{res}} = J H_{J} + J' H_{J'} - r \sum_{i} (c^\dg_{i+\frac{1}{2}, \alpha} f^{\ndg}_{i 1 \alpha} + c^\dg_{i-\frac{1}{2}, \alpha} f^{\ndg}_{i 1 \alpha} + h.c.)$. Notice that the constraint terms do not enter $H_{\txt{res}}$, since the constraints will still be satisfied after the degeneracy of the $f_2$ spinons are lifted.

As usual, from second-order degenerate perturbation theory we obtain an effective Hamiltonian $H^{\txt{eff}}$, given by

\begin{equation} \label{eq:perturbation}
H^{\txt{eff}} = \Proj_{G} H_{\txt{res}} \Proj_{X} \frac{1}{\E_0 - H^{\txt{MF}}} \Proj_{X} H_{\txt{res}} \Proj_{G} \punct{,}
\end{equation}
where $\Proj_{G}$ is the projection operator onto the degenerate ground-state manifold, $\Proj_{X} = 1 - \Proj_{G}$, and $\E_0$ is the unperturbed mean-field ground-state energy.

Since the $f_2$ spinons are decoupled from the electrons and the $f_1$ spinons, the eigenstates $\ket{\psi}$ of $H^{\txt{MF}}$ can be written as product states of the form $\ket{\psi}_{cf_1} \otimes \ket{\psi}_{f_2}$, and correspondingly the Hilbert space decomposes as $\Hilbert = \Hilbert_{cf_1} \otimes \Hilbert_{f_2}$. In this language, the ground-state manifold of $H^{\txt{MF}}$ is $\{ \ket{G}_{cf_1} \} \otimes \Hilbert_{f_2}$, where $\ket{G}_{cf_1}$ is a unique non-degenerate state. Moreover, it can be checked that $H_{\txt{res}}$ factors into the following form:
\begin{equation} \label{eq:Hres_decompose}
H_{\txt{res}} = \left( \sum_{i} \vv{T}_i \cdot f^\dg_{i2\alpha} \frac{\vvsym{\sigma}^{\alpha\beta}}{2} f^{\ndg}_{2i\beta} \right) + \ldots \punct{,}
\end{equation}
where $\vv{T}_i$ and the ``$\ldots$'' operates within $\Hilbert_{cf_1}$. Explicitly,
\begin{align}
\vv{T}_i & = J \left( c^\dg_{i+\frac{1}{2}, \alpha} \frac{\vvsym{\sigma}^{\alpha\beta}}{2} c^{\ndg}_{i+\frac{1}{2}, \beta}  
	+ c^\dg_{i-\frac{1}{2}, \alpha} \frac{\vvsym{\sigma}^{\alpha\beta}}{2} c^{\ndg}_{i-\frac{1}{2}, \beta} \right) \notag \\
	& + J' \left( c^\dg_{i+\frac{1}{2}, \alpha} \frac{\vvsym{\sigma}^{\alpha\beta}}{2} c^{\ndg}_{i-\frac{1}{2}, \beta}  
	+ c^\dg_{i-\frac{1}{2}, \alpha} \frac{\vvsym{\sigma}^{\alpha\beta}}{2} c^{\ndg}_{i+\frac{1}{2}, \beta} \right) 
	\punct{.}
\end{align}

Using Eq.~(\ref{eq:Hres_decompose}), $H^{\txt{eff}}$ becomes
\begin{align} \label{eq:H_eff}
H^{\txt{eff}} & = \sum_{i<j} \J_{ij} \left( f^\dg_{i2\alpha} \frac{\vvsym{\sigma}^{\alpha\beta}}{2} f^{\ndg}_{i2\beta} \right)
	 \cdot \left( f^\dg_{j2\alpha'} \frac{\vvsym{\sigma}^{\alpha'\beta'}}{2} f^{\ndg}_{j2\beta'} \right) + \textrm{const.} \notag \\
	 & = \sum_{i<j} \J_{ij} \vvsym{\s}_i \cdot \vvsym{\s}_j + \txt{const.} \punct{,}
\end{align}
where from translation symmetry it follows that $\J_{ij}$ depends only on $|x_i - x_j|$; i.e., $\J_{ij} = \J_{|i-j|}$. Recognizing $f^\dg_{i2\alpha} \frac{\vvsym{\sigma}^{\alpha\beta}}{2} f^{\ndg}_{i2\beta} \equiv \vvsym{\s}_i$ as a spin-1/2, the second-order degenerate perturbation theory thus maps the $t$-$J$-$J'$ model to an effective spin-1/2 model.

Classically, the ground state of the effective spin model Eq.~(\ref{eq:H_eff}) is a spiral state given by $\vvsym{\s}_{j} = \cos(x_{j}\theta) \uv{n}_1 + \sin(x_{j}\theta) \uv{n}_2 $, where $\uv{n}_1$ and $\uv{n}_2$ are two orthogonal unit vectors and $\theta$ is chosen to minimize the classical energy. However, in 1d, this classical order is expected to be destroyed by quantum fluctuations. 

\subsection{Explicit calculation of \texorpdfstring{$\J_{ij}$}{J\textunderscore ij} and comparison with DMRG results} \label{subsec:J_ij}

Returning to the present case, $\J_{ij}$ in Eq.~(\ref{eq:H_eff}) is given schematically by:
\begin{equation}
\J_{ij} = \left( \sum_{X} \frac{ {}_{cf_1}\bra{G} {T^{z}_{i}}^\dg \ket{X}_{cf_1} \  {}_{cf_1}\bra{X} {T^{z}_{j}}^{\ndg}\!\! \ket{G}_{cf_1} }{\E_{0} - \E_{X}} \right) + c.c. \punct{,}
\end{equation}
in which $\ket{X}_{cf_1}$ are particle-hole excitations from $\ket{G}_{cf_1}$, with $\E_X$ its energy. Note that we have made use of the spin $SU(2)$ symmetry to evaluate $\J_{ij}$ using only the $z$ component of $\vv{T}_i$.

To write down $\J_{ij}$ more explicitly, we Fourier-transform and diagonalize the part of $H^{\txt{MF}}$ that involves only the electron and the $f_1$ spinon, $H^{\txt{MF}}_{cf_1}$, as follows:
\begin{equation}
H^{\txt{MF}}_{cf_1}  = \sum_{k} \sum_{\mu = \pm} \epsilon_{\mu k} \gamma^\dg_{\mu k\alpha} \gamma^{\ndg}_{\mu k\alpha} \punct{,}
\end{equation}
where by convention $\epsilon_{+,k} > \epsilon_{-,k}$. We also define the eigenvectors $u^{a \mu}_k$ via the following:
\begin{equation}
\begin{bmatrix} c_{k\alpha} \\ f_{k\alpha} \end{bmatrix}
= \begin{bmatrix} u^{c+}_{k} & u^{c-}_{k} \\ u^{f+}_{k} & u^{f-}_{k} \end{bmatrix}
\begin{bmatrix} \gamma_{+,k,\alpha} \\ \gamma_{-,k,\alpha} \end{bmatrix} \punct{.}
\end{equation}
Plugging in, we arrive at:
\begin{align} \label{eq:Jij}
\J_{ij} & = \frac{1}{\N^2} \sum_{k,q} \frac{\cos\left( (k-q)(x_j-x_i) \right)}{\epsilon_{-,k} - \epsilon_{+,q}} 
	\left| u^{c-}_k {u^{c+}_q}^{*} \right|^2 \notag \\
	& \times {\left| J (1+e^{iq} e^{-ik}) + J' (e^{iq} + e^{-ik}) \right|}^2 \punct{,}
\end{align}
where $\N$ is the number of unit cells.

Importantly, the $J$ and $J'$ terms in the above equation carry different Fourier phase factors. Thus, even though the ratio $r/t$ in the mean-field ans\"{a}tze depends only on $(J+J')/t$, different effective spin-1/2 models can still be realized for different $J$ and $J'$ having the same sum. 

\begin{figure}
\begin{center}
\subfigure[\label{fig:Classical_Phases_2} Truncated at $\J_2$]{\includegraphics[scale=0.5]{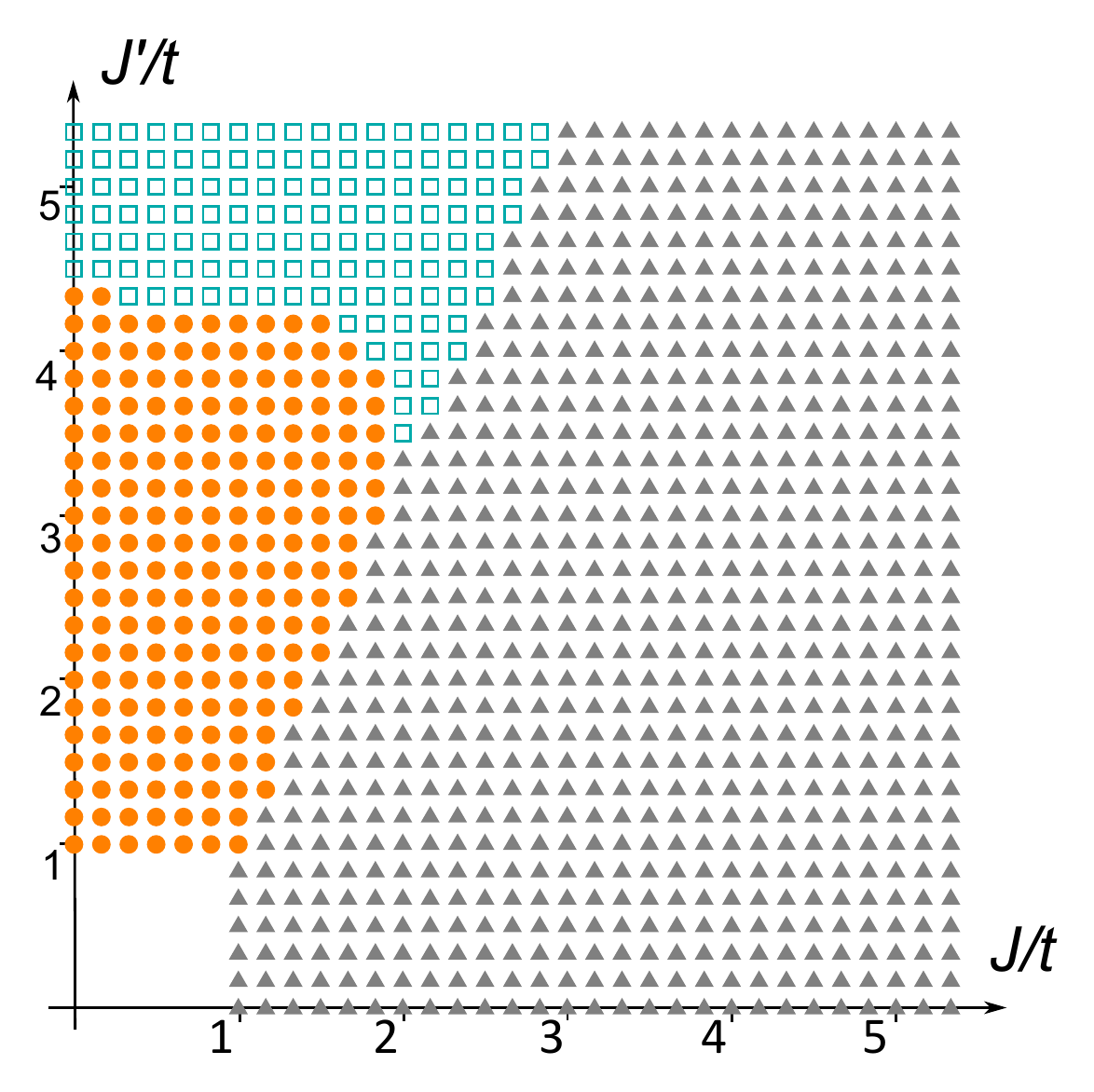}}
\subfigure[\label{fig:Classical_Phases_8} Truncated at $\J_8$]{\includegraphics[scale=0.5]{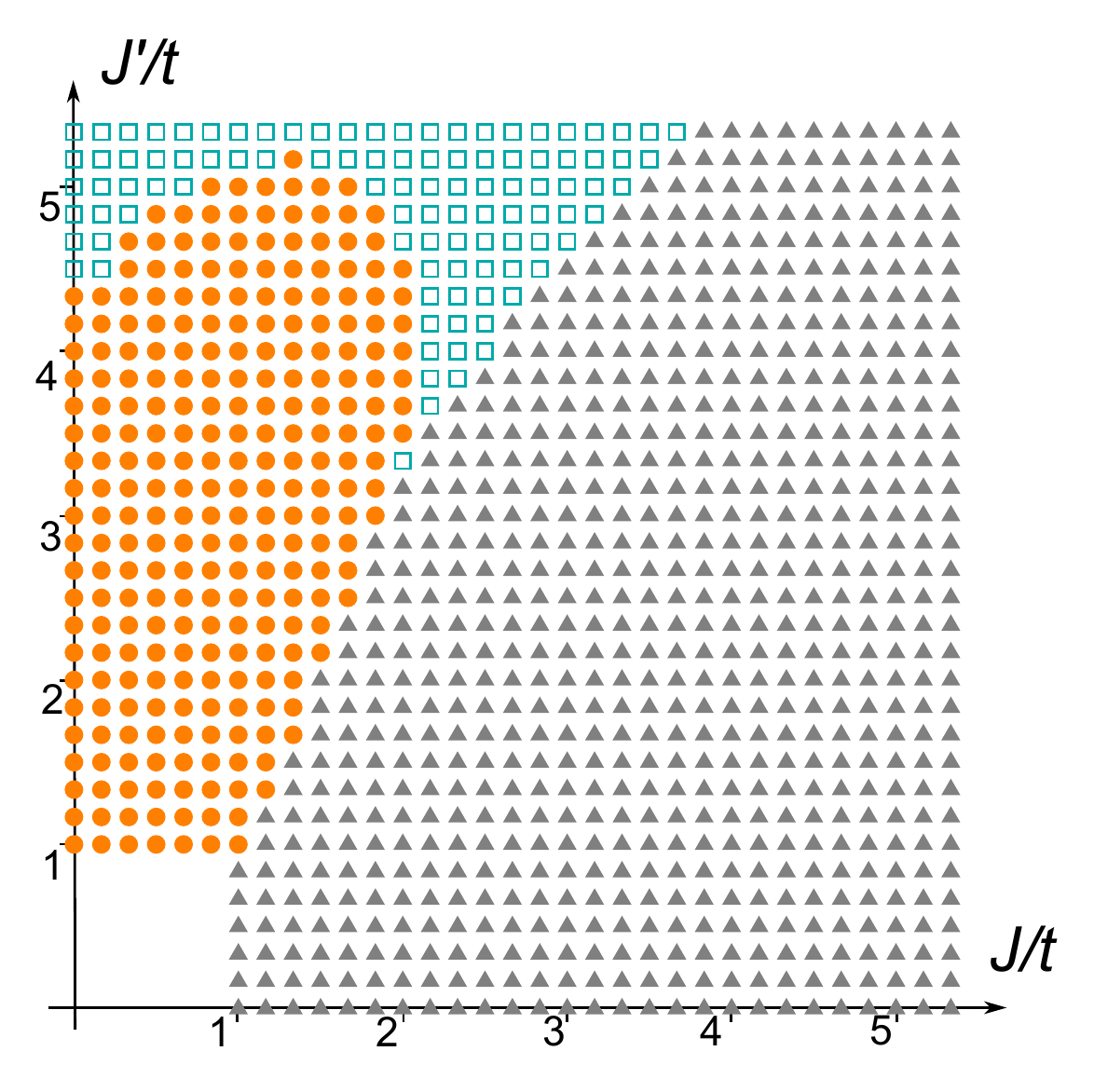}}
\caption{\label{fig:Classical_Phases} (Color online) Classical phase diagram of the effective spin-1/2 model Eq.~(\ref{eq:H_eff}) truncated at (a) $\J_2$ and (b) $\J_8$. Here filled gray triangular symbols indicate the ferromagnetic phase, filled orange circular symbols indicate the antiferromagnetic phase, and unfilled teal square symbols indicate the incommensurate spiral phase. The lower left corner of the phase diagrams are excluded because the gap in $H^{\txt{MF}}$ is too small to accurately calculate $\J_{ij}$.}
\end{center}
\end{figure}

\begin{figure}
\begin{center}
\subfigure[\label{fig:Classical_Angle_4J} $J'=4J$]{\includegraphics[scale=0.3]{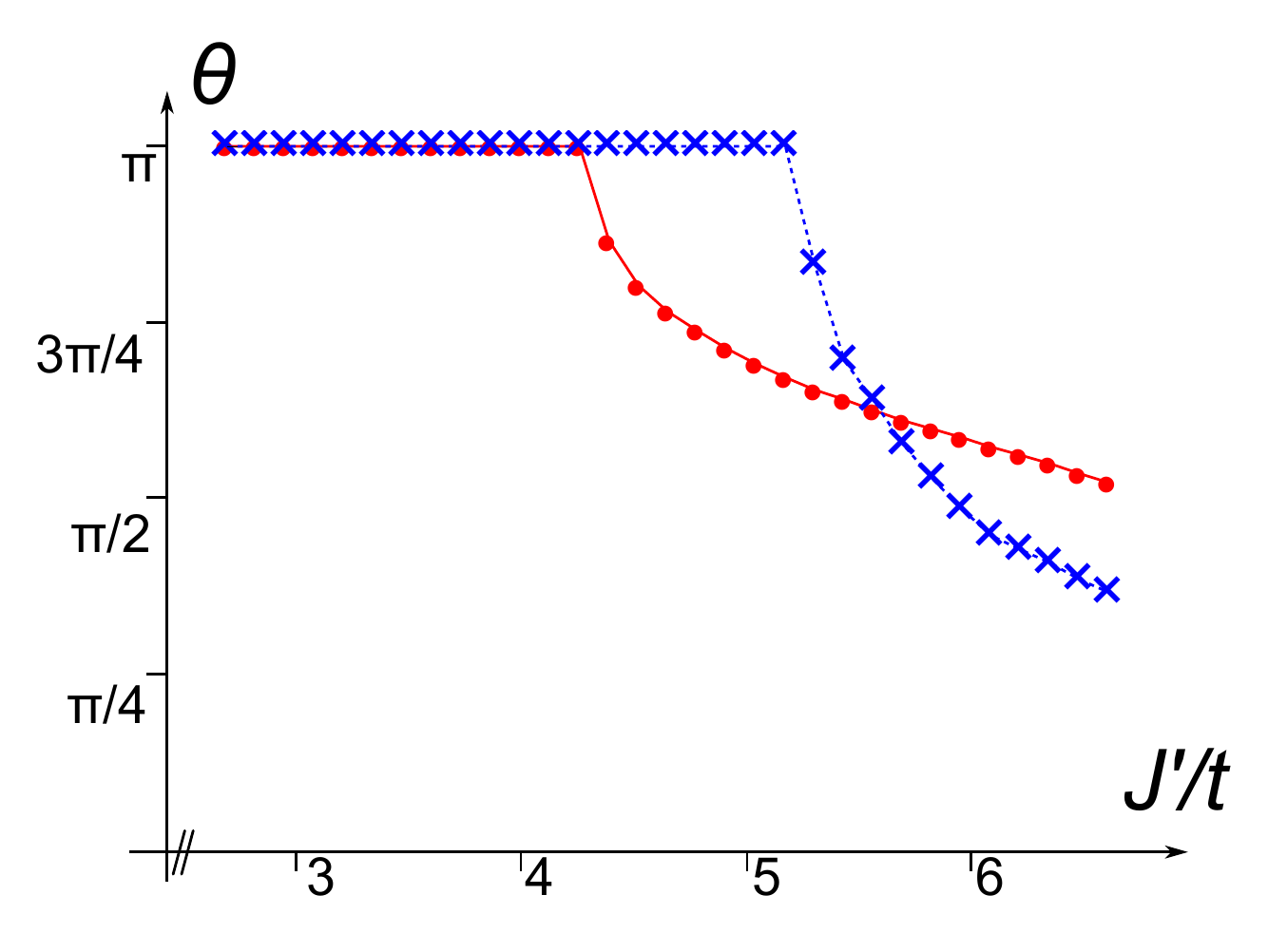}}
\subfigure[\label{fig:Classical_Angle_2J} $J'=2J$]{\includegraphics[scale=0.3]{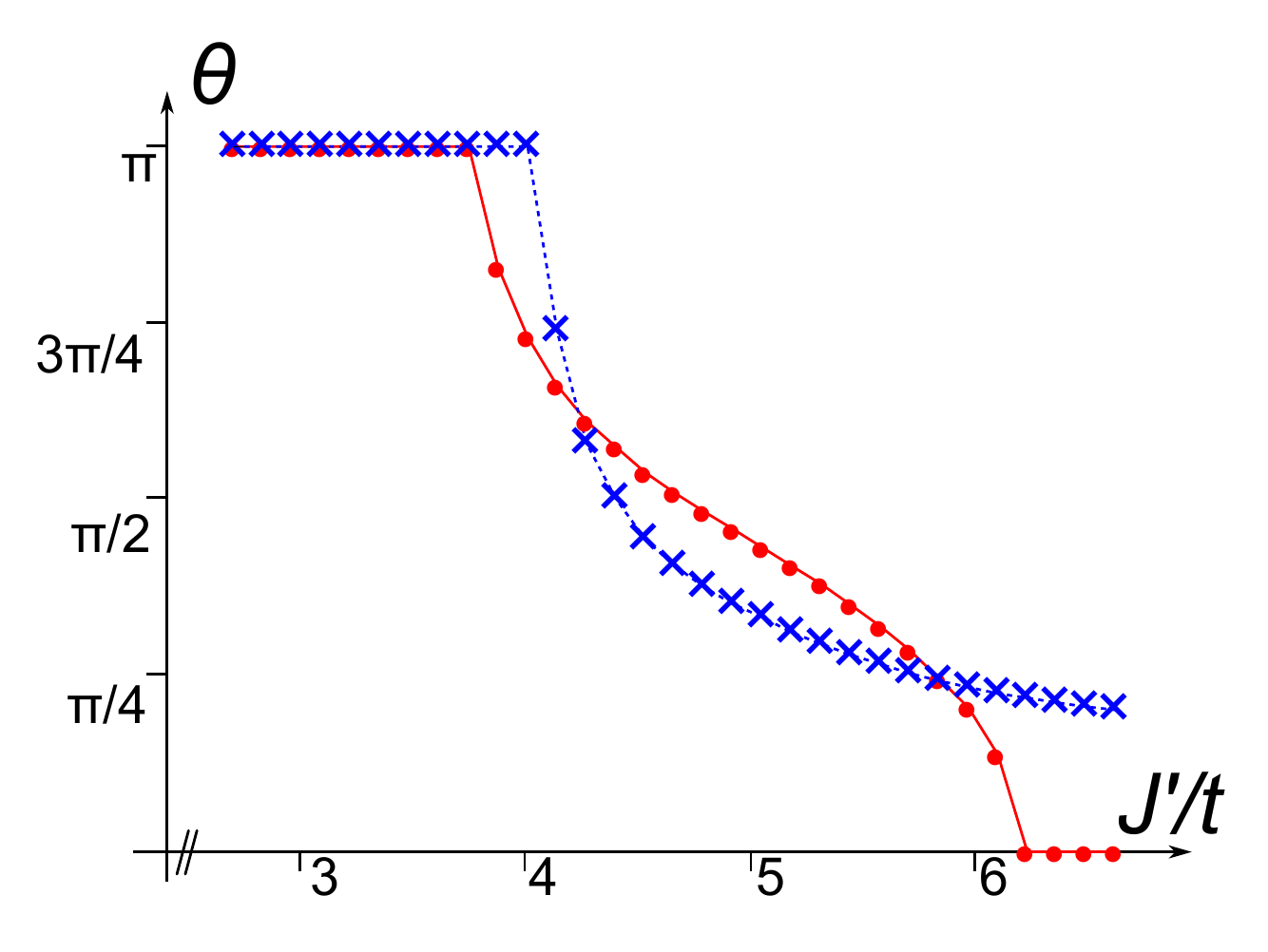}}
\caption{\label{fig:Classical_Angles} (Color online) Classical spiral angle $\theta$ of the effective spin-1/2 model Eq.~(\ref{eq:H_eff}) as function of $J'$ along the line cut (a) $J'=4J$ and (b) $J'=2J$. The red, solid (blue, dashed) curve with filled circle (cross) symbols corresponds to truncation at $\J_2$ ($\J_8$).}
\end{center}
\end{figure}

Calculating $\J_{ij}$ from Eq.~(\ref{eq:Jij}) up to the eighth nearest neighbor, we obtain the classical phase diagrams as shown in Fig.~\ref{fig:Classical_Phases}, in which we distinguish between the antiferromagnetic phase ($\theta = \pi$), the ferromagnetic phase ($\theta = 0$), and the spiral phase ($0 < \theta < \pi$). We also plot the classical spiral angle $\theta$ as function of $J'/t$ along the line cut $J' = 2J$ and $J'=4J$ in Fig.~\ref{fig:Classical_Angles}. 

From these figures it can be seen that even though the details of the phase boundary and the spiral angles are modified as further-neighbor interactions are included, the truncation at $\J_2$ still captures the qualitative aspects of the model reasonably well. Hence, we now focus on the results obtained within this truncation and map the parameters $\J_1$ and $\J_2$ we obtained to the known results from the \emph{quantum} $J_1$-$J_2$ model, in which the ground state is known to exhibit QAF order when $J_1 \gg |J_2| \geq 0$, and undergoes a quantum phase transition into a dimer state at $J_2/J_1 \approx 0.241$. Moreover, as $J_2$ further increases beyond $J_2/J_1 = 1/2$, the peak in the spin-spin correlation originally located at $k = \pi$ also splits into two incommensurate peaks at $k = \pi \pm \delta$, with $\delta$ increasing as $\vartheta = \tan^{-1}(J_2/J_1)$ increases. Eventually, the system becomes ferromagnetic when $\vartheta > \tan^{-1}(-1/4)$.\cite{Okamoto:PLA:1992, Bursill:JPCM:1995, White:PRB:1996, Furukawa:PRB:2012}.

\begin{figure}
\begin{center}
\includegraphics[scale=0.7]{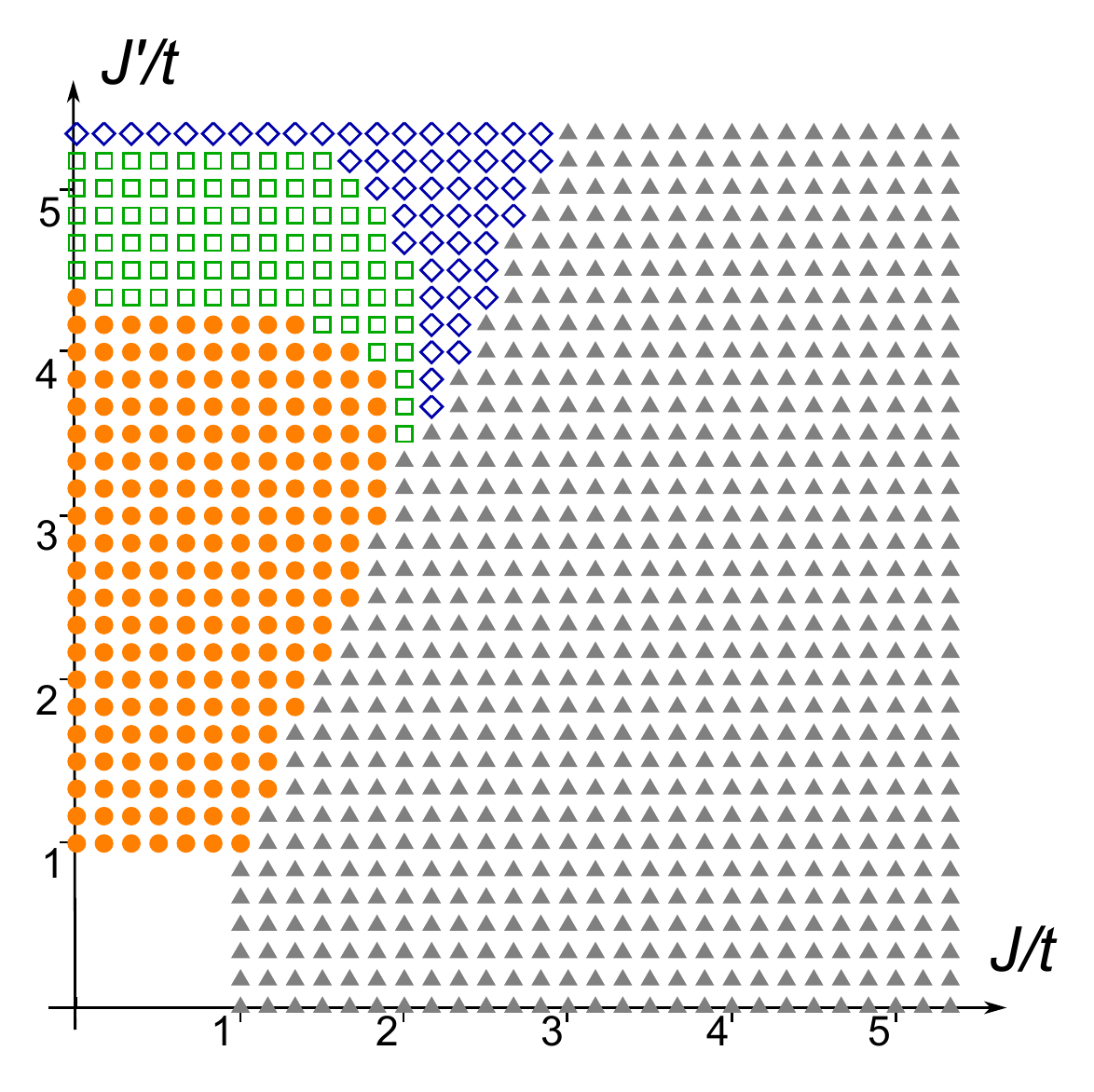}
\caption{\label{fig:J1J2_grid} (Color online) \emph{Quantum} phase diagram of the effective spin-1/2 model Eq.~(\ref{eq:H_eff}), with $\J_{ij}$ computed from Eq.~(\ref{eq:Jij}) and truncated at $\J_2$. Here filled gray triangular symbols indicate the ferromagnetic phase, filled orange circular symbols indicate the \emph{quasi-long-range antiferromagnetic} phase, unfilled green square symbols indicate the \emph{dimer phase} with spin-spin correlation peaked at $k = \pi$, and unfilled violet diamond symbols indicate the \emph{dimer phase} with spin-spin correlation peaked at an \emph{incommensurate} wavevector. The lower left corner of the phase diagram is excluded because the gap in $H^{\txt{MF}}$ is too small to accurately calculate $\J_{ij}$.}
\end{center}
\end{figure}

\begin{figure}
\begin{center}
\includegraphics[scale=0.8]{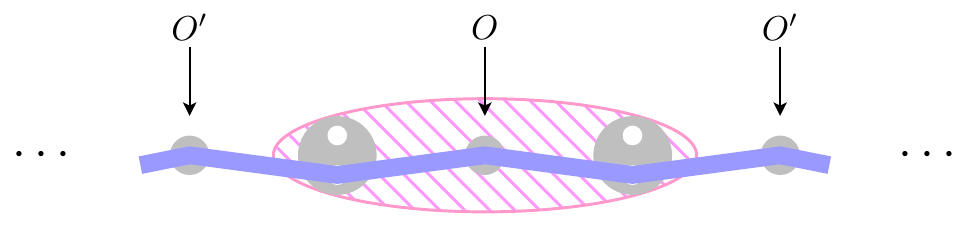}
\caption{\label{fig:Ni-O_dimer} (Color online) Illustration of the two inequivalent oxygen sites when the effective spin-1/2 model is in the dimer phase.}
\end{center}
\end{figure}

The results of our mapping from the $(J/t, J'/t)$ parameter space to the $J_1$-$J_2$ model is shown in  Fig.~\ref{fig:J1J2_grid}. Note that in the present case, there are two inequivalent oxygen sites when the effective spin-1/2 model is in the dimer phase, as illustrated in Fig.~\ref{fig:Ni-O_dimer}. From this, it can be seen that the broken symmetry in the dimer phase of the effective spin-1/2 model is precisely the broken symmetry one would expect from the period-2 CD phase of the DMRG phase diagram (Fig.~\ref{fig:DMRG_phases}). Moreover, even though there is no charge degree of freedom left in the effective spin-1/2 model, charge order is likely to occur when further effects beyond mean-field, e.g.\@ the back-reaction of the dimer order onto the Kondo band insulator formed by the electron and the $f_1$ spinon, are taken into account. Given that the charge deviation in the DMRG CD phase is small, this scenario in which charge order is derived from the ordering of the spin degree of freedom is consistent with the DMRG results. Therefore, we identify the dimer phase of the effective spin-1/2 model with the DMRG CD phase.

Note that in this picture, the charge order in the CD phase is driven by the spin order. Thus, we expect a larger oscillation in spin correlations than in charge density, consistent with the DMRG results (c.f.\@ Figs.~\ref{fig:Ss_DMRG} and \ref{fig:CDW}). We also remark that the amplitude oscillation in $\avg{c^{\dg}_{i\pm1/2} f^{\ndg}_{1i}} \sim \sqrt{\left|\avg{\vv{s}_{i\pm1/2} \cdot \vv{S}_i}\right|}$ cannot be obtained at the mean-field level even if we extend the unit cell to two nickel and two oxygen per cell and allow $r$ to vary from bond to bond, as long as the degeneracy of the $f_2$ spinons is left untouched. This highlights the importance of the dimer formation in the effective spin-1/2 model as the driving mechanism of the oxygen-centered dimer/CD order.

Combining, we see that Fig.~\ref{fig:J1J2_grid} captures the essential aspects of the DMRG phase diagram well, \emph{except for the QS phase}. This is particularly so if one allows for separate renormalizations of $J$ and $J'$ from their bare values, which one can easily imagine to have occurred when various effects that we have neglected are taken into account.

\section{Mean-field identification and effective theory of the QS phase} \label{sec:spiral}

In this section we focus on the QS phase by extending the class of the mean-field ans\"{a}tze considered. Somewhat surprisingly, the flat bands persist even after the class of ans\"{a}tze under consideration is extended (Sec.~\ref{subsec:MF_plus}). To lift such degeneracies without excessive complications, we introduce an additional slave-fermion hopping parameter $t'$, from which we obtain modified mean-field spectra in which for generic parameters a single band crosses the Fermi energy at \emph{two pairs} of Fermi points (Sec.~\ref{subsec:adhoc}). Next we bosonize the effective theory obtained from the low energy fermions (Sec.~\ref{subsec:bosonize}) and search for an appropriate combination of effective interactions that best reproduces the salient features of the QS phase (Sec.~\ref{subsec:interact}). 

Our main result in this section is the identification of the QS phase with a bosonized theory that has one spin and one charge (``C1S1'') mode, in which there is a finite charge gap and the spin field carries an incommensurate wavevector. This interacting bosonized theory is proximate to the spin Bose metal state proposed by Sheng, Motrunich, and Fisher. \cite{DNSheng:PRB:2009} 

\subsection{Extended mean-field ans\"{a}tze} \label{subsec:MF_plus}

Since the class mean-field ans\"{a}tze restricted to $Q_a = Q'_a$ fail to describe the QS phase, we now consider general ans\"{a}tze in which $Q_a$ and $Q'_a$ may be unequal. As in the preceding section, the $U(2)$ gauge redundancy $f_{ia\alpha} \rightarrow U_{i}^{ab} f_{ib\alpha}$ can be used to reduce the number of parameters needed to specify all physically distinct ans\"{a}tze. To begin with, it is clear that we can fix $Q_a = r [1, 0]^T$, with $r \geq 0$. However, such choice does not exhaust the $U(2)$ gauge redundancy as we can still redefine $[ f_{ia1}, f_{ia2} ]^T \rightarrow [ f_{ia1}, e^{i\chi_i} f_{ia2} ]^T$ without changing the form of $Q_a$. This remaining $U(1)$ gauge redundancy allows us to fix the form of $Q'_a$ to be $Q'_a = r' e^{i \phi} [\cos \theta, \sin\theta]^T$, where $r' \geq 0$, $\phi \in [0, 2\pi)$ and $\theta \in [0, \pi)$.

In addition, we may further demand the mean-field Hamiltonian to be time-reversal invariant, which fixes $\phi = 0$ or $\pi$ in the above expression for $Q'_a$. Absorbing the sign coming from $\phi = \pi$ into $\theta$, it thus suffices to take $Q_a = r [1, 0]^T$ and $Q'_a = r' [\cos \theta, \sin\theta]^T$, with $r,r' \geq 0$ and $\theta \in [0, 2\pi)$.

\begin{figure}
\begin{center}
\includegraphics[scale=0.6]{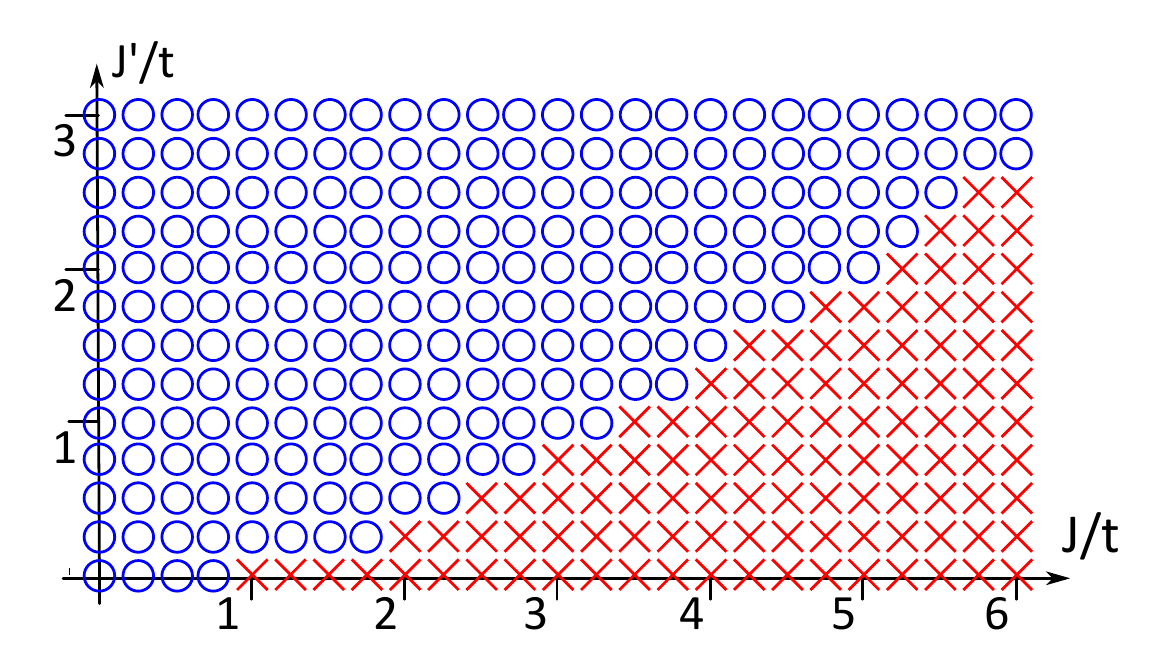}
\caption{\label{fig:Unequal_grid} (Color online) Results from energy minimization for the general mean-field ans\"{a}tze. Here unfilled blue circle (red cross) indicates states for which $ \sqrt{ \sum_{a} \left| \avg{ f^\dg_{ia\alpha} c_{i+\frac{1}{2}\alpha} } - \avg{ f^\dg_{ia\alpha} c_{i-\frac{1}{2}\alpha} } \right|^2 }$ $\leq$ ($>$) $0.05$. }
\end{center}
\end{figure}

In Fig.~\ref{fig:Unequal_grid} we plot the results obtained from minimizing $\avg{H_{tJJ'}}_{\txt{MF}}$ with respect to the general ans\"{a}tze parametrized above. From the figure it can be seen that states with $Q_a \neq Q'_a$ emerge only for $J \gtrsim 2 J'$. Comparing Fig.~\ref{fig:Unequal_grid} with Figs.~\ref{fig:Classical_Phases} and \ref{fig:J1J2_grid}, it can be seen that the region for which $Q_a \neq Q'_a$ is deep inside the ferromagnetic phase of the effective spin-1/2 model and does not fit well to the location of the QS phase in the DMRG phase diagram Fig.~\ref{fig:DMRG_phases}.

However, it should be noted that the low-energy gauge structure of an ansatz with $\theta \neq 0$ is markedly different from that of an ansatz with $\theta = 0$. Specifically, when $\theta = 0$, the gauge transformation $[ f_{ia1}, f_{ia2} ]^T \rightarrow [ f_{ia1}, e^{i\chi_i} f_{ia2} ]^T$ leaves the mean-field ansatz invariant, implying that a $U(1)$ gauge field remains gapless in the low-energy effective theory. In contrast, when $\theta \neq 0$ there is no continuous transformation that leaves the ansatz invariant, hence no gapless gauge field remains in the low-energy effective theory. Because of this difference, it is conceivable that the energies of mean-field ans\"{a}tze with $\theta \neq 0$ may renormalize differently from those with $\theta = 0$, thus opening the possibility that ans\"{a}tze with $\theta \neq 0$ may become favorable in the region of parameter space that corresponds to the QS phase. In what follows, we shall stop worrying about the mean-field energetics and instead focus on whether the low-energy effective theory obtained from mean-field ans\"{a}tze with $\theta \neq 0$ can account for the QS phase.

In the Schwinger fermion decomposition Eq.~(\ref{eq:Schwinger}), $\avg{ \vv{S}_i \cdot \vv{s}_{i\pm\frac{1}{2}} } \propto \sum_{a} \left|\avg{f^\dg_{ia\alpha} c^{\ndg}_{i\pm\frac{1}{2}\alpha}} \right|^2$. Thus, setting $r \neq r'$ in the mean-field ansatz will result in a mean-field state for which $\avg{ \vv{S}_i \cdot \vv{s}_{i+\frac{1}{2} } } \neq \avg{ \vv{S}_i \cdot \vv{s}_{i-\frac{1}{2} } }$. However, from DMRG we know that within numerical accuracy $\avg{ \vv{S}_i \cdot \vv{s}_{i+\frac{1}{2} } } = \avg{ \vv{S}_i \cdot \vv{s}_{i-\frac{1}{2} } }$ (c.f.\@ Fig.~\ref{fig:Ss_DMRG}). Hence, in the remaining we shall consider only the case in which $r = r'$ but $\theta \neq 0$.

\begin{figure}
\begin{center}
\subfigure[\label{fig:admixplot_1} $t'=0$, $\theta = 0$]{\includegraphics[scale=0.36]{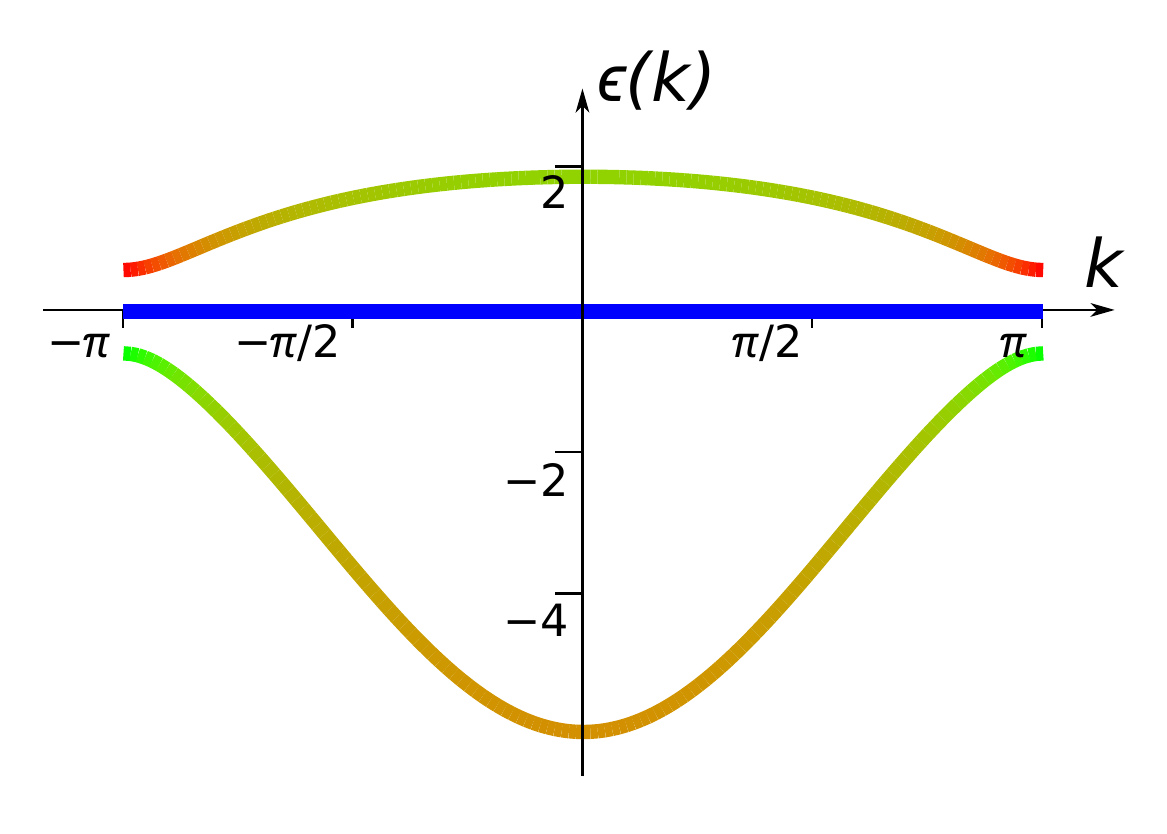}}
\subfigure[\label{fig:admixplot_2} $t'=0$, $\theta = \pi/8$]{\includegraphics[scale=0.36]{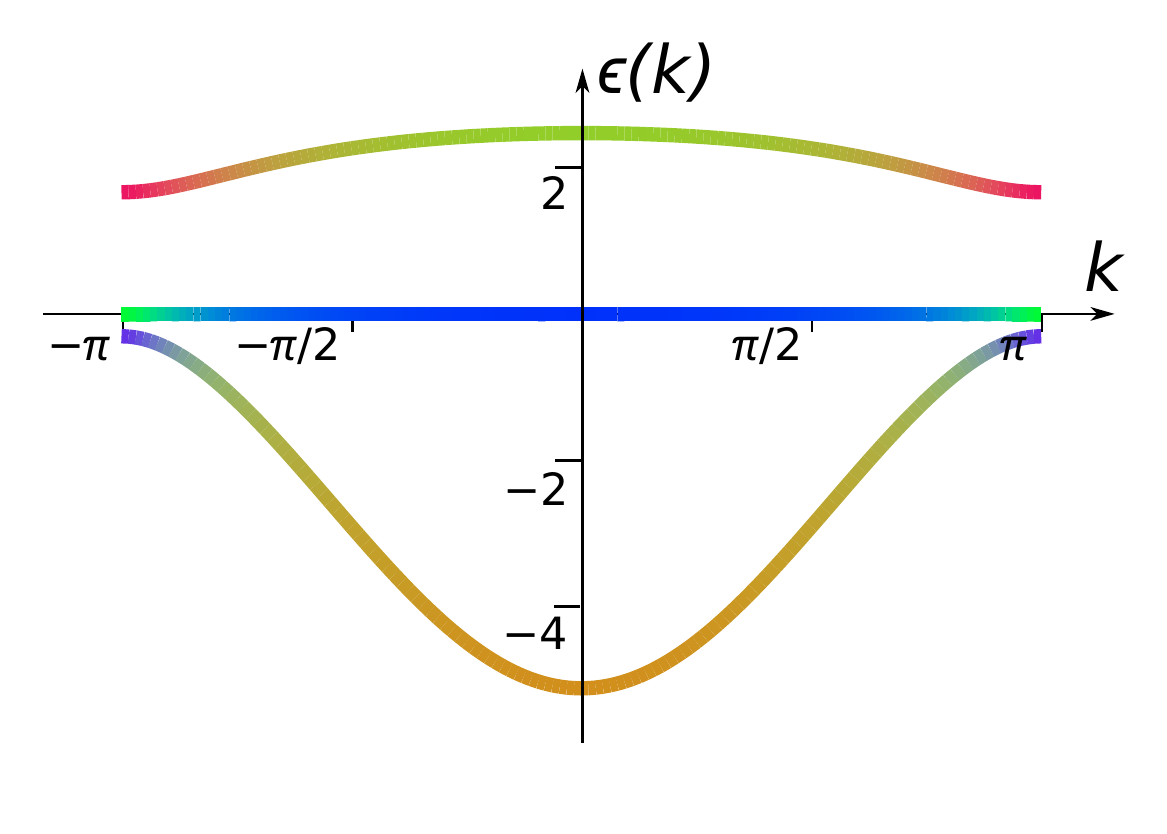}}
\subfigure[\label{fig:admixplot_3} $t'=-0.1 t$, $\theta = \pi/8$]{\includegraphics[scale=0.36]{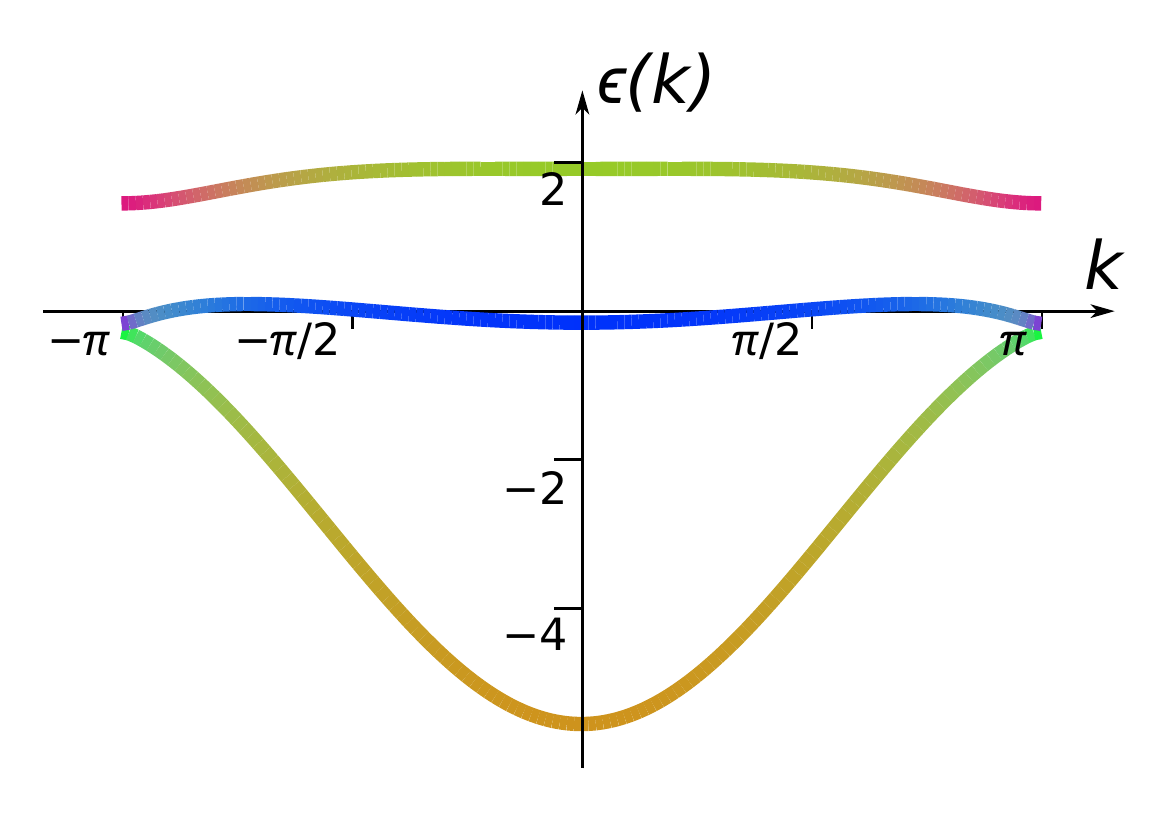}}
\subfigure[\label{fig:admix_color} Color scheme]{\includegraphics[scale=0.24]{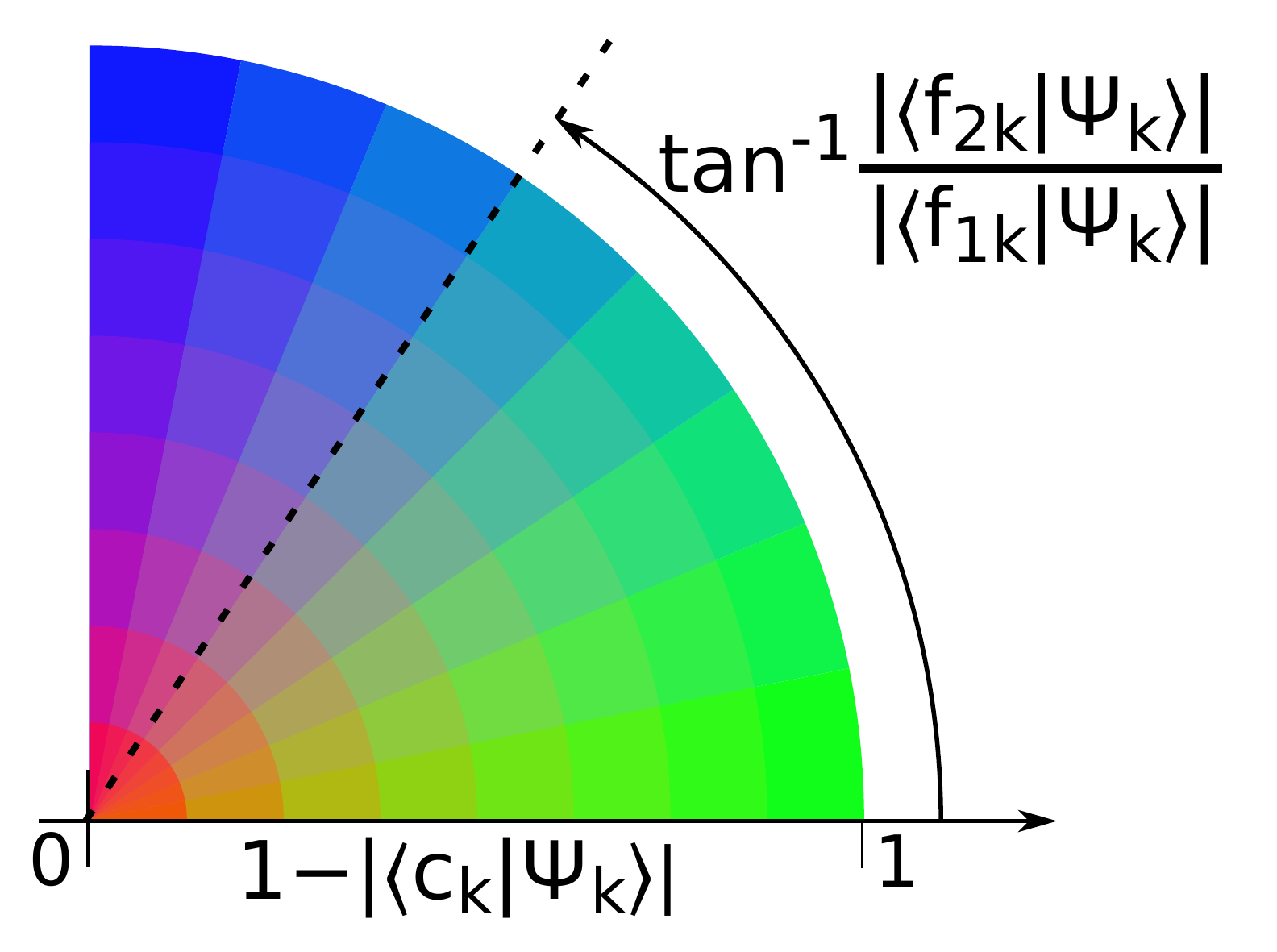}}
\subfigure[\label{fig:admixplot_3a} $t'=-0.1 t$, $\theta = \pi/8$, details of the second band]{\includegraphics[scale=0.6]{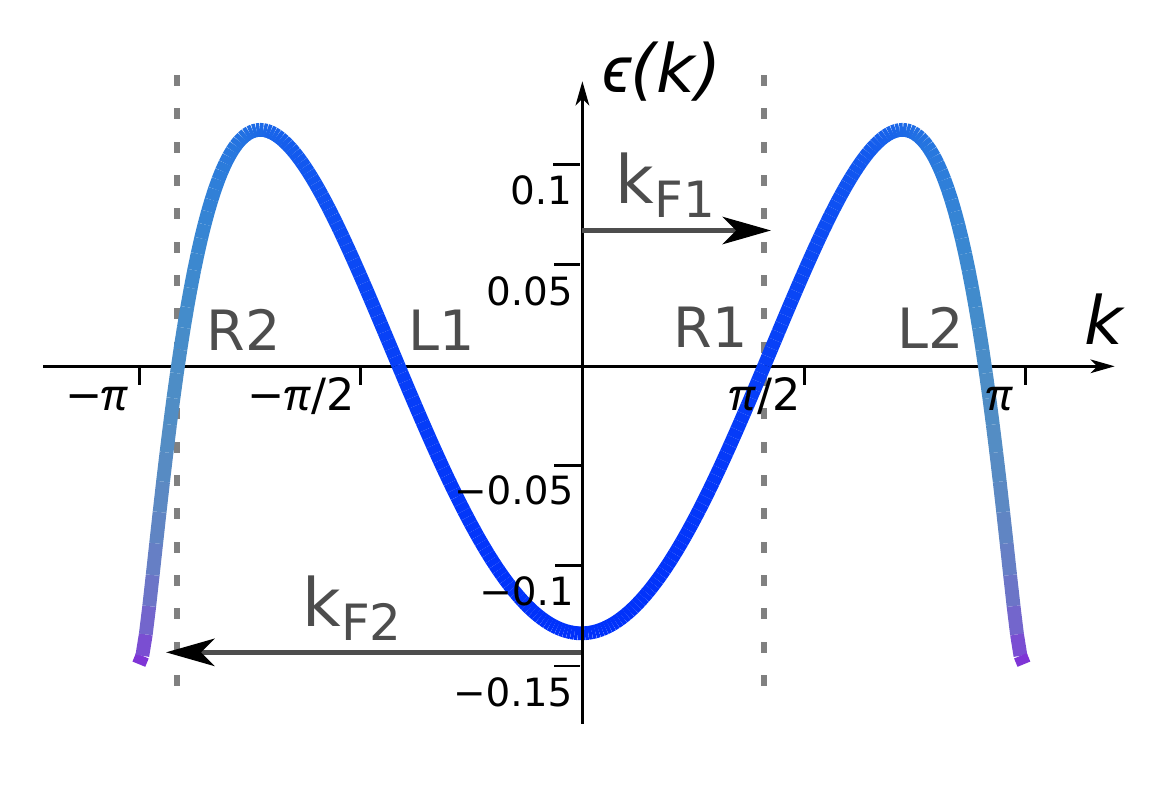}}
\caption{\label{fig:admixplots} (Color online) Band structures obtained from $H^{\txt{MF}}$ with the additional $f$-hopping term $t'$, for $r/t = r'/t = 1.812$ and various values of $t'$ and $\theta$. The colors encode the eigenvector composition of the bands, with red $\sim$ $c_{k}$, green $\sim$ $f_{1k}$, and blue $\sim$ $f_{2k}$ [the detailed color scheme is shown in panel~(d)]. Panel~(e) also defines the convention for the bosonization treatment.}
\end{center}
\end{figure}

For illustration, we pick $r/t = 1.812$, which when $\theta = 0$ corresponds to, e.g., $J = 8/3$ and $J' =16/3$, from our previous calculation (a region which from Fig.~\ref{fig:J1J2_grid} one might expect to be proximate to the QS phase), and introduce an \textit{ad hoc} value of $\theta = \pi/8$ to the mean-field ansatz. The original mean-field spectrum with $\theta = 0$ and the modified spectrum with $\theta = \pi/8$ are shown in Figs.~\ref{fig:admixplot_1} and \ref{fig:admixplot_2}, respectively. Somewhat surprisingly, the mean-field spectrum is largely unaffected by the change of $\theta$. In particular, the flat band at Fermi energy continues to appear in the spectrum.\footnote{Technically, the constraints Eq.~(\ref{eq:constraints}) and the half-filled condition in such case are solved by assuming that each zero-energy state carries a ``democratic'' weight of 1/2. i.e., $\avg{n_k} = 1/2$ of all states in the flat band.} This behavior seems to be a generic feature for this class of ans\"{a}tze. i.e., this flat band exists for general values of $\theta$ and $r/t$. However, importantly, the compositions of the eigenstates in this flat band are modified, as can be inferred from the coloring of the band in Fig.~\ref{fig:admixplot_2} (color online), which shows that the states near $k=\pm\pi$ have large wavefunction overlaps with the $f_1$ spinons.  

\subsection{Lifting the degeneracies in the class of extended mean-field ans\"{a}tze} \label{subsec:adhoc}

As before, effects beyond mean field are expected to lift the degeneracy of the flat band. In principle, one can apply degenerate perturbation theory as presented in the preceding section, \emph{but with two significant modifications}: First, the zero-energy mean-field single-particle state $\gamma_{i0\alpha}$ on site $i$ now has to be constructed from Wannier orbitals. Since both species of spinons have non-zero Kondo hoppings, $\gamma_{i0\alpha}$ is no longer locally conserved. Consequently, terms of the form $\gamma^\dg_{i0\alpha} \gamma^{\ndg}_{j0\alpha}$ ($i \neq j$) can appear in the effective Hamiltonian. Second, the constraints enforcing terms $\lambda$ and $\vv{N}$ in $H^{\txt{MF}}$ now depend crucially on the precise manner in which the degeneracy is lifted, and hence cannot be left out in the residual interaction $H_{\txt{res}}$. On the technical level, it is challenging to perform the perturbative calculation with the two modifications stated above.

In addition, on the conceptual level, in the QS state singularities appear in \emph{both} the nickel and electron spin-spin correlations, the latter of which are absent in the other phases in the DMRG phase diagram. Such singularities in the electron spin-spin correlation cannot be easily captured in the perturbation theory, since the Wannier orbitals $\gamma_{i0\alpha}$ are predominately spinon in character. 

Therefore, here we take an alternative approach in which an additional \textit{ad hoc} spinon-spinon hopping term $t' \sum_{i} \big( f^\dg_{i,a,\alpha} f^{\ndg}_{i+1,a,\alpha} + h.c. \big)$ is introduced into the mean-field Hamiltonian $H^{\txt{MF}}$. Such term can be thought of as arising from the spinon mean-field decomposition of the nearest neighbor spin-spin interaction term $\J_{ij} \vvsym{\s}_{i} \cdot \vvsym{\s}_{j}$ generated by the degenerate perturbation.

In Figs.~\ref{fig:admixplot_3} and \ref{fig:admixplot_3a} we plot the resulting mean-field spectrum for $t' =  - 0.1 t < 0$, with $r/t = 1.812$ and $\theta = \pi/8$ as before, from which we see that the resulting spectrum now has four Fermi crossings at incommensurate wavevectors. This four-crossing spectrum appears to be a general feature of the mean-field ans\"{a}tze when $\theta \neq 0$ and $t' \neq 0$; i.e., they exist as long as $\theta \neq 0$ and $t'$ is small but non-zero. However, the Fermi velocities at the four crossing points will be inverted when $t' > 0$. More importantly, the resulting crossings will be predominately spinon in character, while for $t' < 0$ the crossings at $k = \pm k_{F2}$ near $\pm \pi$ will have a non-negligible electron weight [for the parameters used in Figs.~\ref{fig:admixplot_3} and \ref{fig:admixplot_3a}, $\left| \avg{c_{k_{F2}\alpha}|\gamma_{k_{F2}0\alpha}} \right| \approx 0.3$]. We shall therefore take $t' < 0$ and consider the bosonized theory of the archetypal band structure shown in Fig.~\ref{fig:admixplot_3a}. Apart from the minor complication arising from the matrix elements resulting from the compositions of the low-energy fermions in terms of the original ($c$,$f_1$,$f_2$) fermions, the bosonized theory of the four-Fermi-crossings band structure shown in Fig.~\ref{fig:admixplot_3a} has been studied extensively by Sheng, Motrunich, and Fisher in the context of the so-called spin Bose metal (SBM).\cite{DNSheng:PRB:2009} In their construction, a two-leg triangular strip (zigzag chain) is considered, in which a four-site ring exchange term $K$ is added to the $J_1$-$J_2$ Heisenberg model. In that model, the SBM phase, characterized in part by singularities at incommensurate wavevectors in various correlation functions, is observed for a range of $J_2/J_1$ when $K$ is sufficiently large (at minimum $K/J_1 \gtrsim 0.2$).

Here, we shall adopt most of their notations and keep our account down to the essentials by referring our readers to Ref.~\onlinecite{DNSheng:PRB:2009} for details.

\subsection{Bosonization of the mean-field theory} \label{subsec:bosonize}

As in Ref.~\onlinecite{DNSheng:PRB:2009}, we define eight species of low-energy fermions $\psi_{Pa\alpha}$, one for each Fermi point, in which $P = R/L \equiv +/-$ labels the two propagation directions, $\alpha = \up, \dn$ labels the two spins, and $a = 1,2$ corresponds the two Fermi wavevectors $k_{Fa}$. The two Fermi wavevectors are chosen such that fermions at $k_{Fa}$ are right-moving, and that $|k_{F2}| > |k_{F1}|$ [see Fig.~\ref{fig:admixplot_3a} for illustration]. Note that the Fermi wavevectors satisfy the relation $k_{F1} + k_{F2} = -\pi/2$.

Next, we bosonize the fermions as follows:
\begin{equation} \label{eq:bosonization}
\psi_{Pa\alpha} \propto \eta_{a\alpha} e^{i(\varphi_{a\alpha} + P \theta_{a\alpha} ) } \punct{,}
\end{equation}
in which $\varphi_{a\alpha}$ and  $\theta_{a\alpha}$ are bosonic fields that satisfy $\left[\varphi_{a\alpha}(x), \varphi_{b\beta}(x') \right] = \left[\theta_{a\alpha}(x), \theta_{b\beta}(x') \right] = 0$ and $\left[\varphi_{a\alpha}(x), \theta_{b\beta}(x') \right] = i\pi \delta_{ab} \delta_{\alpha\beta} \Theta(x-x')$ (here $\Theta(x)$ is the Heaviside step function with regularization $\Theta(0) = 1/2$), while $\eta_{a\alpha}$ are the Klein factors satisfying $\{ \eta_{a\alpha},  \eta_{b\beta} \} = 2 \delta_{ab} \delta_{\alpha\beta}$.

As in Ref.~\onlinecite{DNSheng:PRB:2009}, in addition to the above $\{1\up, 1\dn, 2\up, 2\dn\}$ basis for the bosonized fields, it is useful to introduce also the $\{1\rho, 1\sigma, 2\rho, 2\sigma\}$ basis and $\{\rho+, \rho-, \sigma+, \sigma-\}$ basis, defined by the following canonical transformations of the $\theta$ fields and $\varphi$ fields (the transformations for the $\varphi$ fields are given by replacing every $\theta$ with $\varphi$ in the equations below):
\begin{align}
\theta_{a\rho} & = \frac{\theta_{a\up} + \theta_{a\dn}}{\sqrt{2}} \punct{,} 
	& \theta_{a\sigma} & = \frac{\theta_{a\up} - \theta_{a\dn}}{\sqrt{2}} 
	& & (a=1,2) \punct{;}  \\
\theta_{\mu+} & = \frac{\theta_{1\mu} + \theta_{2\mu}}{\sqrt{2}} \punct{,} 
	& \theta_{\mu-} & = \frac{\theta_{1\mu} - \theta_{2\mu}}{\sqrt{2}} 
	& & (\mu = \rho, \sigma) \punct{.} 
\end{align}
As customary, we shall refer to fields with index $\rho$ as the charge fields and fields with index $\sigma$ as the spin fields.

At the level of fermion bilinears, the wavevectors $\pm 2 k_{Fa}$ $(a =1,2)$, $\pm \pi/2$, and $\pm (k_{F1} - k_{F2})$ are brought out, and the corresponding bosonized expressions for the nickel spin $\vv{S}_{k}$ and the electron density $\delta n_{k}$ are given by
\begin{align}
S^{x}_{2k_{Fa}} & \propto e^{\sqrt{2} i \theta_{a\rho}} \sin(\sqrt{2}\varphi_{a\sigma}) \punct{,} \label{eq:bilinear_start}\\ 
S^{y}_{2k_{Fa}} & \propto e^{\sqrt{2} i \theta_{a\rho}} \cos(\sqrt{2}\varphi_{a\sigma}) \punct{,} \\
S^{z}_{2k_{Fa}} & \propto e^{\sqrt{2} i \theta_{a\rho}} \sin(\sqrt{2}\theta_{a\sigma}) \punct{,} \\
\delta n_{2k_{Fa}} & \propto e^{\sqrt{2} i \theta_{a\rho}} \cos(\sqrt{2}\theta_{a\sigma}) \punct{;}
\end{align}

\begin{align}
S^{x}_{\pi/2} & \propto 
	\omitted e^{-i\theta_{\rho+}} e^{i\theta_{\sigma-}} \sin(\varphi_{\rho-} - \varphi_{\sigma+}) \notag \\
	& + \omitted e^{-i\theta_{\rho+}} e^{-i\theta_{\sigma-}} \sin(\varphi_{\rho-} + \varphi_{\sigma+})
	\punct{,} \\
S^{y}_{\pi/2} & \propto 
	\omitted e^{-i\theta_{\rho+}} e^{i\theta_{\sigma-}} \cos(\varphi_{\rho-} - \varphi_{\sigma+}) \notag \\
	& + \omitted e^{-i\theta_{\rho+}} e^{-i\theta_{\sigma-}} \cos(\varphi_{\rho-} + \varphi_{\sigma+})
	\punct{,} \\
S^{z}_{\pi/2} & \propto 
	\omitted e^{-i\theta_{\rho+}} e^{i\theta_{\sigma+}} \sin(\varphi_{\rho-} - \varphi_{\sigma-}) \notag \\
	& + \omitted e^{-i\theta_{\rho+}} e^{-i\theta_{\sigma+}} \sin(\varphi_{\rho-} + \varphi_{\sigma-})
	\punct{,} \\
\delta n_{\pi/2} & \propto 
	\omitted e^{-i\theta_{\rho+}} e^{i\theta_{\sigma+}} \sin(\varphi_{\rho-} - \varphi_{\sigma-}) \notag \\
	& + \omitted e^{-i\theta_{\rho+}} e^{-i\theta_{\sigma+}} \sin(\varphi_{\rho-} + \varphi_{\sigma-}) \punct{;}
\end{align}

\begin{align}
S^{x}_{k_{F1} - k_{F2}} & \propto 
	\omitted e^{i\theta_{\rho-}} e^{-i\theta_{\sigma+}} \sin(\varphi_{\rho-} - \varphi_{\sigma+}) \notag \\
	& + \omitted e^{i\theta_{\rho-}} e^{i\theta_{\sigma+}} \sin(\varphi_{\rho-} + \varphi_{\sigma+})
	\punct{,} \\
S^{y}_{k_{F1} - k_{F2}} & \propto 
	\omitted e^{i\theta_{\rho-}} e^{-i\theta_{\sigma+}} \cos(\varphi_{\rho-} - \varphi_{\sigma+}) \notag \\
	& + \omitted e^{i\theta_{\rho-}} e^{i\theta_{\sigma+}} \cos(\varphi_{\rho-} + \varphi_{\sigma+})
	\punct{,} \\
S^{z}_{k_{F1} - k_{F2}} & \propto 
	\omitted e^{i\theta_{\rho-}} e^{-i\theta_{\sigma-}} \sin(\varphi_{\rho-} - \varphi_{\sigma-}) \notag \\
	& + \omitted e^{i\theta_{\rho+}} e^{i\theta_{\sigma-}} \sin(\varphi_{\rho-} + \varphi_{\sigma-})
	\punct{,} \\
\delta n_{k_{F1} - k_{F2}} & \propto 
	\omitted e^{i\theta_{\rho+}} e^{-i\theta_{\sigma-}} \sin(\varphi_{\rho-} - \varphi_{\sigma-}) \notag \\
	& + \omitted e^{i\theta_{\rho+}} e^{i\theta_{\sigma-}} \sin(\varphi_{\rho-} + \varphi_{\sigma-}) \punct{.} \label{eq:bilinear_end}
\end{align}
where $\omitted$ represents various numerical and \emph{Klein} factors, which are not important for our purposes. Also, the bosonized expressions for electron spin $\vv{s}_k$ are essentially the same as that of the nickel spin except for changes in the numerical factors in the $(\ldots)$ due to matrix elements. As usual, $\mathcal{O}^{\ndg}_{-k} = \mathcal{O}_{k}^\dg$ for $\vv{S}_{k}$, $\vv{s}_{k}$, and $\delta n_{k}$.

In the absence of any residual interactions, the Lagrangian density for the bosonized fields is given by:
\begin{equation} \label{eq:bosonized_Lg}
\Lg_0 = \frac{1}{2\pi} \sum_{a\alpha} \left( \frac{1}{v_a} (\pd_\tau \theta_{a\alpha})^2 + v_a (\pd_x \theta_{a\alpha})^2 \right) \punct{.}
\end{equation}
It is important to remark that in our case the bosonized fields are the \emph{only} low-energy degree of freedom remaining in the theory. More precisely, recall that $\lambda$ and $\vv{N}$ should properly be thought of as fluctuating fictitious gauge fields, and that $Q_a$ and $Q'_a$ should properly be thought of as fluctuating bosonic fields. However, for the mean-field ans\"{a}tze that we now consider, all fictitious gauge fields have been gapped through the Anderson--Higgs mechanism, with various transverse components of $Q_a$ and $Q'_a$ fluctuations serving as the corresponding Goldstone boson that are ``eaten up.'' The remaining fluctuations of $Q_a$ and $Q'_a$ are gapped, upon integrating out high-energy degrees of freedom if not at the bare level. Therefore, contrary to Ref.~\onlinecite{DNSheng:PRB:2009}, there is no \emph{a priori} reason for $\theta_{\rho+}$ to be pinned. 

\subsection{Interactions in the bosonized theory} \label{subsec:interact}

In the absence of any pinnings of the bosonic fields, the low-energy effective theory described by Eq.~(\ref{eq:bosonized_Lg}) is a $c=4$ Luttinger liquid. However, as the DMRG results show a charge gap in QS phase, we shall accept as an empirical matter that $\theta_{\rho+}$ is pinned, which can happen if the eight-fermion interaction $\psi^\dg_{R1\up} \psi^\dg_{R1\dn} \psi^\dg_{R2\up} \psi^\dg_{R1\dn} \psi^{\ndg}_{L1\up} \psi^{\ndg}_{L1\dn} \psi^{\ndg}_{L2\up} \psi^{\ndg}_{L2\dn} + h.c. \propto \cos(4\theta_{\rho+})$ is sufficiently strong. Moreover,  since there are \emph{two} incommensurate Fermi wavevectors $k_{F1}$ and $k_{F2}$, the correlation functions in such theory are expected to exhibit singularities at \emph{multiple} incommensurate wavevectors that are integer combinations of $k_{F1}$ and $k_{F2}$, as well as at wavevectors that are integer multiples of $\pi/2 = -(k_{F1} - k_{F2})$. Since the correlation functions obtained in DMRG (Fig.~\ref{fig:correlations}) show prominent singularities at merely \emph{one} pair of incommensurate wavevectors $k = \pi \pm \delta$, the free theory given by Eq.~(\ref{eq:bosonized_Lg}) seems to be inconsistent with DMRG. Such inconsistency could in principle be accounted for if all the undesired singularities are suppressed by non universal amplitudes. However, it is more natural to consider scenarios in which some of the bosonic fields are pinned by interaction.

As explained in Ref.~\onlinecite{DNSheng:PRB:2009}, assuming that chiral interactions lead only to velocity renormalizations, the four-fermion interactions schematically consists of three parts, $\Lg_{\txt{int-4}} = W + V_{\rho} + V_{\sigma}$. In terms of the bosonized fields, these read:
\begin{align}
W &= \cos(2\varphi_{\rho-}) \bigg[ 
	4 w^{\rho}_{12} \Big( \cos(2\varphi_{\sigma-}) - \KleinProd \cos(2 \theta_{\sigma-}) \Big) \notag \\
	& - w^\sigma_{12} \Big( \cos(2\varphi_{\sigma-}) + \KleinProd \cos(2 \theta_{\sigma-}) + 2 \KleinProd \cos(2 \theta_{\sigma+}) \Big) \bigg] \punct{,} \\
V_{\rho} & = \sum_{a} \frac{\lambda^{\rho}_{aa}}{2\pi^2} \Big( (\pd_{x} \theta_{a\rho})^2 - (\pd_{x} \varphi_{a\rho})^2 \Big) \notag \\
	& + \frac{\lambda^{\rho}_{12}}{\pi^2} \Big( (\pd_{x} \theta_{1\rho}) (\pd_{x} \theta_{2\rho}) - (\pd_{x} \varphi_{1\rho}) (\pd_{x} \varphi_{2\rho}) \Big) \punct{,}\\
V_{\sigma} & = \sum_{a} \lambda^{\sigma}_{aa} \cos(2 \sqrt{2} \theta_{a\sigma}) + 2\lambda^{\sigma}_{12} \KleinProd \cos(2\theta_{\sigma+}) \cos(2\varphi_{\sigma-}) \notag \\
	& +\sum_{a} \frac{\lambda^{\sigma}_{aa}}{8\pi^2} \Big( (\pd_{x} \varphi_{a\sigma})^2 - (\pd_{x} \theta_{a\sigma})^2\Big) \notag \\
	& + \frac{\lambda^{\sigma}_{12}}{4 \pi^2} \Big( (\pd_{x} \varphi_{1\sigma}) (\pd_{x} \varphi_{2\sigma}) - (\pd_{x} \theta_{1\sigma}) (\pd_{x} \theta_{2\sigma}) \Big) \punct{,}
\end{align}
where $w^{\mu}_{12}$ and $\lambda^{\mu}_{ab}$ ($a,b = 1,2$ and $a \leq b$; $\mu = \rho, \sigma$) are parameters that control the interaction strength. Moreover, as in Ref.~\onlinecite{DNSheng:PRB:2009}, the parameters $\lambda^{\sigma}_{ab}$ satisfy the following RG equations at the one-loop level:
\begin{equation} \label{eq:lambda_RG}
\frac{d\lambda^\sigma_{aa}}{d\ell} = - \frac{(\lambda^\sigma_{aa})^2}{2\pi v_a} \punct{,} \qquad  
	\frac{d\lambda^\sigma_{12}}{d\ell} = - \frac{(\lambda^\sigma_{12})^2}{\pi (v_1 + v_2)} \punct{,}
\end{equation}
such that the only instabilities caused by $V_{\sigma}$ arise from $\lambda^{\sigma}_{ab} < 0$.

Note that even after incorporating $V_\rho$, the resulting Lagrangian density $\Lg' = \Lg_0 + V_\rho$ remains quadratic. However, scaling dimensions of operators that contain the charge fields will be modified. As a result, the terms in $W$ will in general acquire scaling dimensions that are different from their bare value. Hence, we can consider separately the case in which $W$ contains the most relevant terms and the case in which $V_{\sigma}$ contains the most relevant terms.

If $W$ contains the most relevant terms, then in general $\varphi_{\rho-}$ is pinned, and correspondingly $\theta_{\rho-}$ is completely disordered. From the bosonization formula and from the definition of $\theta_{\rho-}$ it can be check that the bosonized expression of a product of fermions does not contain $\theta_{\rho-}$ if and only if it carries momenta that are multiples of $\pi/2$. Consequently, the pinning of $\varphi_{\rho-}$ will kill the singularities of any correlation functions located at incommensurate wavevectors. This is inconsistent with the DMRG results. Hence, we conclude that in the bosonized theory of the QS phase $W$ must be irrelevant.

Next we consider the case in which $V_\sigma$ contains marginally relevant terms. In the simplest scenarios, the terms with which $\lambda^{\sigma}_{ab} < 0$ contain mutually commuting variables. There are four such cases: 
\begin{enumerate}[(i)]
\item $\lambda^\sigma_{11} > 0$, $\lambda^\sigma_{22} > 0$, $\lambda^\sigma_{12} < 0$ ; \label{item:lambda_i}
\item $\lambda^\sigma_{11} < 0$, $\lambda^\sigma_{22} < 0$, $\lambda^\sigma_{12} > 0$ ; \label{item:lambda_ii}
\item $\lambda^\sigma_{11} < 0$, $\lambda^\sigma_{22} > 0$, $\lambda^\sigma_{12} > 0$ ; \label{item:lambda_iii}
\item $\lambda^\sigma_{11} > 0$, $\lambda^\sigma_{22} < 0$, $\lambda^\sigma_{12} > 0$ . \label{item:lambda_iv}
\end{enumerate}
In cases (\ref{item:lambda_i}) and (\ref{item:lambda_ii}) two mutually
commuting spin fields are pinned. Consequently, all spin-spin
correlations must be void of singularities. Thus, these cases are
inconsistent with the DMRG results. In cases (\ref{item:lambda_iii}) and
(\ref{item:lambda_iv}), one spin field associated with wavevector
$k_{Fa}$ is pinned. Thus, at the level of fermion bilinears, the only
singularities in the spin-spin correlations that survive are the ones at
$\pm 2 k_{Fa'}$, where $a' \neq a$
[c.f. Eqs.~(\ref{eq:bilinear_start})--(\ref{eq:bilinear_end})]. These
singularities can be identified with the singularities present in DMRG
QS phase. Moreover, recall that in the DMRG results there are broad
peaks in the nickel (but not electron) spin-spin correlation located
roughly at wavevectors $\pi \pm k_{\txt{sing}}$, where $\pm
k_{\txt{sing}}$ are the wavevectors of which the prominent singularities
are seen. These broad peaks can be interpreted as the remnants of the
singularities at $\pm 2 k_{Fa}$ after the associated spin field is
pinned. Recall that the Fermi points at $\pm k_{F2}$ have non-negligible
electron character while the Fermi points at $\pm k_{F1}$ are
predominately spinon in character. Since in the DMRG the singularities are observed in both the nickel and the electron spin-spin correlation, while the broad peaks are observed only in the nickel spin-spin correlation, we may identify the state obtained in DMRG with scenario (\ref{item:lambda_iv}). Since we also assume that $\theta_{\rho+}$ is pinned, the resulting state would be a $c=2$ Luttinger liquid with one charge mode and one spin mode (``C1S1''). 

One potential objection to this identification is that it implies that the electron density-density correlation function $\avg{\delta n_{-k} \delta n_{k} }$  possesses singularities at $\pm 2 k_{F1}$ and $\pm 2 k_{F2}$, which are not observed. However, it is known that when the charge fluctuation is reintroduced to the SBM, as the system approaches the Mott transition and when it is in the insulating phase, the non-universal amplitudes in the density-density correlation can be sufficiently small that in numerics it can appear smooth.\cite{mishmash:_metal} The situation that we found in the DMRG study of the present model may correspond to such a situation.

More generally, one might pose the question of whether the DMRG results can be explained by other combinations of pinned fields, which can arise from higher-order interactions. Here we briefly consider such possibilities.

Since we assume that $\theta_{\rho+}$ is pinned, the $\{ \rho+$, $\rho- \}$ basis is the appropriate basis to describe the charge sector. The only question left for the charge sector is whether $\varphi_{\rho-}$ or $\theta_{\rho-}$ can be pinned. As already mentioned, the pinning of $\varphi_{\rho-}$ would kill all correlations at incommensurate wavevectors and hence is inconsistent with the DMRG results. As for $\theta_{\rho-}$, one can check that it carries an incommensurate momentum $k_{F1} - k_{F2}$ (i.e., $ \theta_{\rho-} \rightarrow \theta_{\rho-} + k_{F1} - k_{F2}$ under the translation $i \rightarrow i+1$), and hence cannot be pinned. Thus, unless one appeals to vanishingly small non-universal amplitudes, exactly one spin degree of freedom must be gapped to produce the single pair of prominent singularities in the spin-spin correlation observed in DMRG. Therefore, we are left with a ``C1S1'' case similar to the one we analyzed, except that the pinned spin field may more generally be a linear combination of the $1\sigma$ and $2\sigma$ fields.

\section{Discussion and Conclusions} \label{sec:conclude}

In this manuscript we considered a 1d underscreened Kondo chain with alternating spin-1 and electron sites, which in addition to the familiar electron hopping and Kondo term also contained a spin-dependent hopping term. We analyzed the model numerically using DMRG and found that the phase diagram consists of a ferromagnetic (F) phase, a quasi-long-range antiferromagnetic (QAF) phase, a charge-density (CD) ordered phase, and, importantly, a quasi-long-range spiral (QS) phase, in which singularities in the spin-spin correlation are observed at incommensurate wavevectors. The phases in our model can in principle be distinguished experimentally by many experimental means.  For example, the charge order can be observed by x-ray scattering, or by its effect on the crystal structure.  The magnetic order is readily probed by neutron scattering, susceptibility measurements, and nuclear magnetic resonance.   

To interpret the DMRG results, we introduced a slave-particle representation of the spin-1 spins, from which we obtained a mean-field Hamiltonian. Taking the simplest class of mean-field ans\"{a}tze and making use of second-order degenerate perturbation theory, we were able to map the mean-field Hamiltonian to an effective spin-1/2 model, from which follows a quantum mean-field phase diagram that resembles the one obtained from DMRG, with the exception of the QS phase. We then focused on the QS phase, by extending our class of mean-field ans\"{a}tze and considering the bosonized interacting theory that arose from them. By considering various possible interactions at the four-fermion level, we argued that the QS phase is best reproduced by an interacting bosonized theory with one charge and one spin degree of freedom (``C1S1''), in which the spin fields carry an incommensurate wavevector. This particular interacting bosonized theory is proximate to the spin Bose metal phase proposed by Sheng, Motrunich, and Fisher\cite{DNSheng:PRB:2009}. Consequently, our results point to a possible route to obtain gapless quantum spin liquids with singularities at incommensurate wavevectors that does not involve the ``ring-exchange'' terms, \cite{DNSheng:PRB:2009, Mishmash:PRB:2011} and might pave the way for realizing such states in 1d as well as constructing similar states in higher dimensions.

While we believe that the current work presents a coherent and comprehensive analysis of the system we considered, we also mention a few opportunities for further study. On the numerical side, it is clearly desirable to obtain the central charge of the QS phase from DMRG and check against the $c=2$ prediction from mean-field theory.  This requires either further computation resources or improvements of the methodology.   Moreover, it would be helpful to check the \emph{projected} energetics of the mean-field ans\"{a}tze we proposed, and compare them against the energy obtained from DMRG, via variation Monte Carlo (VMC). Unfortunately, because the slave-particle representation we used requires additional constraints on top of the local conservation of spinon number [c.f.\@ Eq.~\ref{eq:constraints}], such a calculation is challenging.  On the analytic side, it would be desirable to develop a method that can \emph{directly} obtain the CD phase from the slave-particle representation without appealing to an indirectly argument based on symmetry.  We also note that there is an alternative slave-particle representation of the spin-1 spins in the literature,\cite{ZXLiu:PRB:2010, Bieri:PRB:2012} and it would be good to check whether results similar to those we obtained can be derived in this alternative slave-particle representation.   

Finally, we return to the nickel valence controversy that motivated this work, which in the simplest case can be considered as a three-dimensional (3d) extension of the present model.   From our mean-field analysis and from general arguments on Kondo problems we are quite convinced that the low-energy physics of such 3d model will still be described by an effective spin-1/2 model, which tends to exhibit static spin orders.   However, the dimer phase in the effective spin-1/2 model, which we argue to correspond to the oxygen-centered charge-density order, is a special feature in 1d, and is unlikely to persist to 3d.   In any case, our DMRG phase diagram shows no sign of nickel-centered charge order (or its symmetry-equivalent order---we do not insist upon any significant charge accumulation), which has been interpreted as due to local Kondo singlet formation.   Thus, the direct extension of our model to 3d is unlikely to exhibit any charge order, whether it is oxygen-centered or nickel-centered.   In our opinion this points to the necessity of including explicit coupling to oxygen motions in the mechanism of charge ordering.  It would be interesting to explore this directly by explicitly including lattice degrees of freedom in a model similar to ours.  We leave this topic for future study.

\begin{acknowledgments}

We acknowledge useful discussions with George Sawatzky, Andy Millis, Susanne Stemmer, Jim Allen, Patrick Lee, Matthew Fisher, and Olexei Motrunich.  This work was supported by the ARO through Grant No.~W911-NF-09-1-0398 (H.-C. J.), the NSERC of Canada (J.G.R.), the MRSEC program of the NSF under Award No. DMR-1121053 (L.B.), and the KITP's NSF Grant, PHY-11-25915 (W.-H.K. and L.B.). We also acknowledge support from the Center for Scientific Computing at the CNSI and MRL: an NSF MRSEC (DMR-1121053) and NSF CNS-0960316.

\end{acknowledgments}



\end{document}